\documentclass[onecolumn]{aastex63}

\usepackage{lineno}
\usepackage{subfiles, amssymb, amsmath, url}
\usepackage{bm}
\usepackage[T1]{fontenc}

\newcommand{\bfI}{\bm{I}}
\newcommand{\bfC}{\bm{C}}
\newcommand{\bfR}{\bm{R}}
\newcommand{\bfW}{\bm{W}}
\newcommand{\Rmat}{\bm{R}}
\newcommand{\bfE}{\bm{E}}
\newcommand{\bfF}{\bm{F}}

\newcommand{\bfk}{\boldsymbol{k}}
\newcommand{\bfx}{\boldsymbol{x}}
\newcommand{\bfr}{\bm{r}}

\newcommand{\bfdelta}{\boldsymbol{\delta}}
\newcommand{\bfg}{\boldsymbol{g}}

\DeclareMathOperator{\sinc}{sinc}

\submitjournal{ApJ}

\shorttitle{Pencil Beam-IM Cross Correlation}
\shortauthors{Visbal \& McQuinn}

\graphicspath{{./}{figures/}}

\renewcommand{\vec}[1]{\boldsymbol{\mathrm{#1}}}

\begin{document}

\title{Cross Correlation of Pencil-Beam Galaxy Surveys and Line-Intensity Maps: An Application of the James Webb Space Telescope}

\author{Eli Visbal}
\affiliation{Department of Physics and Astronomy and Ritter Astrophysical Research Center, University of Toledo, 2801 W. Bancroft Street, Toledo, OH, 43606, USA}

\author{Matthew McQuinn}
\affiliation{University of Washington, Department of Astronomy, 3910 15th Ave NE, Seattle, WA, 98195, USA}

\begin{abstract}
Line-intensity mapping (IM) experiments seek to perform statistical measurements of large-scale structure with spectral lines such as 21cm, CO, and Lyman-$\alpha$ (Ly$\alpha$). A challenge in these observations is to ensure that astrophysical foregrounds, such as galactic synchrotron emission in 21cm measurements, are properly removed. One method that has the potential to reduce  foreground contamination is to cross correlate with a galaxy survey that overlaps with the IM volume. However, telescopes sensitive to high-redshift galaxies typically have small field of views (FOVs) compared to IM surveys. Thus, a galaxy survey for cross correlation would necessarily consist of pencil beams which sparsely fill the IM volume. In this paper, we develop the formalism to forecast the sensitivity of cross correlations between IM experiments and pencil-beam galaxy surveys. We find that a random distribution of pencil beams leads to very similar overall sensitivity as a lattice spaced across the IM survey and derive a simple formula for random configurations that agrees with the Fisher matrix formalism.
We explore examples of combining high-redshift \emph{James Webb Space Telescope} (\emph{JWST}) observations with both a \emph{SPHEREx}-like Ly$\alpha$ IM survey and a 21cm experiment based on the \emph{Hydrogen Epoch of Reionization Array (HERA)}. We find that the \emph{JWST}-\emph{SPHEREx}  case is promising, leading to a total signal-to-noise of ${\sim}5$ after 100 total hours of \emph{JWST} (at $z=7$). We find that \emph{HERA} is not well-suited for this approach owing to its drift-scan strategy, but that a similar experiment that can integrate down on one field could be.
\end{abstract}


\section{Introduction} 
\label{sec:intro}

Some of the most exciting reionization-era signals will be measured using surveys that span large swaths of the sky.  This includes IM efforts observing 21cm radiation such as \emph{MWA}, \emph{LOFAR}, \emph{HERA} and \emph{SKA} \citep{refId0,DeBoer2017, koopmans, 2022JATIS...8a1007B}.  It also includes IM with lines such Ly$\alpha$, H$\alpha$, CO, and [CII] with instruments like \emph{SPHEREx} \citep{2014arXiv1412.4872D}, \emph{CDIM} \citep{2019BAAS...51g..23C}, \emph{FYST} \citep{2022A&A...659A..12K}, \emph{COMAP} \citep{2022ApJ...933..182C}, \emph{TIME} \citep{2021ApJ...915...33S}, and \emph{CONCERTO} \citep{2020A&A...642A..60C}.  These intensity maps will often contain strong foregrounds (e.g., galactic synchrotron emission in the 21cm case) and imperfect removal of these foregrounds could masquerade as signal.  

Cross correlations with a tracer of the high-redshift universe would be the most robust way to mitigate foregrounds.  Previous studies have investigated potential cross correlations of intensity maps with the CMB \citep{2010MNRAS.402.2617T, 2013ApJ...779..124M} and wide-field narrow band Ly$\alpha$ emitter surveys \citep{2009ApJ...690..252L,2016MNRAS.459.2741S, 2018MNRAS.479.2754K, 2020MNRAS.492.4952V, 2022MNRAS.512..792C}. The signal-to-noise (S/N) in such cross correlations is often found to be small because the line-of-sight oriented structures IM efforts target are orthogonal to the sky-plane structures these other surveys are generally most sensitive to.  One promising idea is to correlate the Subaru HyperSuprimeCam narrow band Ly$\alpha$ emitter survey at $z=6.6$ with a \emph{LOFAR} 21cm intensity map.  Forecasts are that this could provide a detectable signal with $\text{S/N}\sim 2-4$ \citep{2020MNRAS.492.4952V}. Another previously explored idea is cross correlating IM surveys with other IM surveys that map distinct lines \citep{2010JCAP...11..016V, 2011ApJ...741...70L, 2011ApJ...730L..30C}.

Spectroscopic galaxy surveys in the optical/near infrared provide excellent line-of-sight resolution and so are a natural match for the high line-of-sight resolution of IM surveys and, hence, for cross correlation.  Unfortunately, obtaining spectroscopic redshifts for high-redshift sources is challenging.  A promising high-redshift spectroscopic catalogue could come from the \emph{Roman} Space Telescope's slitless spectrograph.  Predictions for the cross correlations of a dedicated survey with this instrument and the \emph{HERA} intensity map have found ${\rm S/N} \sim 10$ \citep{2022arXiv220509770L}.  The slitless spectroscopy of \emph{Roman} will not be as sensitive as spectra from the largest ground based optical telescopes or \emph{JWST}. However, in contrast to Roman, the small field of view of these telescopes is poorly matched to the wide fields of many IM surveys.    

Here we consider how feasible it would be to use surveys with narrow fields to detect cross correlations with wide fields. Namely, we consider whether cross correlations with a large number of pencil beams sampling across an IM survey could yield a sufficient sensitivity to be useful for measuring or confirming signals (i.e.~to ensure proper removal of spurious foregrounds).  Calculating the S/N of such a survey is complicated by the non-continuous survey geometry.  We develop the framework to do this and forecast the S/N for correlating \emph{JWST} with \emph{SPHEREx} and \emph{HERA}.  This study is most related to \citet{2015ApJ...800..128B}, who found a potentially detectable correlation of the galaxy counts within a \emph{JWST} field and the pixel intensity of \emph{MWA} and \emph{HERA} maps at that location. It is also related to previous work considering cross correlating between the Ly$\alpha$ forest and low spectral resolution IM survey \citep{2021MNRAS.501.3883R}.

This paper is organized as follows.  Section~\ref{sec:formalism} presents the formalism for cross correlations with pencil beams, reducing complex expressions to simple formulas in the limit of random pointings. Section~\ref{sec:JWST} discusses the sensitivity specifications for \emph{JWST} spectroscopy.  The sensitivity of  cross correlations are presented in Sections~\ref{sec:sensitivity} and \ref{sec:hera}, where we first consider an instructive noiseless case and, then, we consider the cases of \emph{SPHEREx} Ly$\alpha$ and \emph{HERA} 21cm measurements.  We finish with concluding thoughts.  Throughout, we assume a $\Lambda {\rm CDM}$ cosmology with parameters consistent with \cite{2014A&A...571A..16P}: $\Omega_{\rm m} = 0.32$, $\Omega_{\Lambda} = 0.68$, $\Omega_{\rm b} = 0.049$, $h=0.67$, $\sigma_8=0.83$, and $n_{\rm s} = 0.96$.

\section{Cross-power spectrum sensitivity formalism}
\label{sec:formalism}
In this section, we describe the formalism to estimate the cross-power spectrum sensitivity for an IM survey cross correlated with a galaxy survey comprised of pencil beams. The goal of the cross correlation could be either to remove foregrounds or to confirm an IM power spectrum measurement that could be contaminated with astrophysical foregrounds. Thus, we only consider information in the cross-correlation (i.e. we assume the auto-correlations of the signal do not contribute to the S/N). Here we include an outline of the formalism and the most important resulting equations. Additional details can be found in Appendices~\ref{ap:fourier}, \ref{ap:covariances}, \ref{sec:crossestim} and \ref{sec:signtonoisedetection}.

We begin by defining a data vector for the IM survey, $\bfI$, which includes (as separate components) both the real and imaginary parts of each Fourier mode sampled by the IM survey. Thus, there are two components corresponding to the $i$'th mode, 
${\rm Re}(\tilde{{\delta}}_{\rm I}(\bfk_i))$ and ${\rm Im}(\tilde{\delta}_{\rm I}(\bfk_i))$. The quantity $\tilde{{\delta}}_{\rm I}$ is defined as the Fourier transform of the spatially fluctuating intensity of the IM signal after subtracting off the mean intensity. As discussed below, the number of modes sampled, $N_{k}$, depends on the size and spatial resolution of the IM survey and the length of $\bfI$ is $2\, N_{k}$. 
We only include modes with positive values of the wavevector component parallel to the line of sight, $k_{\parallel}$, since the IM measurements being purely real quantities makes these modes redundant with those having negative values of this component (because~$\tilde{\delta}_{\rm I}(\bfk) = \tilde{\delta}_{\rm I}(-\bfk)^* $).

While a logical basis of measurements for IM surveys is Fourier modes, for galaxy pencil beams a more natural basis is the galaxy overdensity in each pointing's field at a given line-of-sight wavenumber. Thus, we take the components of our galaxy survey data vector for a combined set of pencil beams to be the real and imaginary parts of 

\begin{equation}
g_{i,j} = \int d^2 \bfx' \hat{\delta}_{\rm g}(\bfx', k_{{\rm \parallel},i}) W_{\rm g} \left (\bfx_{j} - \bfx' \right ),
\label{eqn:deltaG}
\end{equation}
where $\hat{\delta}_{\rm g}(\bfx', k_{{\parallel},i})$ is the partial Fourier transform of galaxy overdensity (transformed only in the line-of-sight direction; see Appendix~\ref{ap:fourier}), $j$ indexes the different pencil beams, $W_{\rm g}$ is the galaxy survey window function for one pointing, and $\bfx_{j}$ is the (2D) location on the sky of the center of one pencil beam. Here $\bfx'$ represents a 2D position on the sky and $k_{{\parallel},i}$ is the wavenumber of the mode in the line-of-sight direction. The number of such modes is set by the line-of-sight spatial coverage and resolution. 

The window function $W_{\rm g}$ is defined to be zero outside of the pencil-beam FOV and constant within and normalized so that $\int d^2 \bfx' W_{\rm g}(\bfx') = 1 $.  This integral over the window function means that we are not retaining information on clustering on angular scales smaller than the survey field.  This approximation is justified for narrow pencil beams with instruments such as \emph{JWST} when cross correlated with IM observations that do not resolve angular scales smaller than the pencil beam FOV.  For galaxy surveys with wider-field instruments, like \emph{Roman}, this approximation is less justified.

The standard estimate for the minimum error of a set of parameters, ${p}_i$, (for instance values of the cross-power spectrum in different $k$-bins) is given by the Fisher matrix

\begin{equation}
F_{ij} = \frac{1}{2}{\rm Tr} \left [ \bfC^{-1}\bfC_{,i}     \bfC^{-1}\bfC_{,j}  \right ],
\label{eqn:fisher}
\end{equation}
where $\bfC_{,i} \equiv \frac{\partial \bfC}{\partial p_i}$ \citep{1997ApJ...480...22T}.
Here $\bfC$ is the covariance matrix of the data vector $\boldsymbol{d}$ (which in our case would include the components of $\bfI$, as well as $\bfg$ including each pencil beam). Bounds on the error of $\widehat{p}_i$ -- here the hat indicates an estimated quantity -- are given by $\sigma_i \gtrsim \sqrt{[\bm{F}^{-1}]_{i,i}}$, with measurements in cosmology often saturating this bound because of the Gaussianity of cosmological signals. 

In our application, the traditional Fisher matrix expression (Eq.~\ref{eqn:fisher}) would use all of  the information from both the IM and galaxy surveys.  As mentioned above, this would not estimate the sensitivity of most interest, as we want to \emph{only} use the information in cross correlation (since our aim is to mitigate foregrounds through cross correlation or to confirm an auto-power spectrum measurement).  Appendix~\ref{ap:estimator} presents a derivation of the optimal quadratic estimator that only uses cross-power information (extending the work of \citealt{2018PhRvD..98j3526V}). In this case, and assuming the parameters indexed by $p_i$, are the cross power bandpowers (i.e., the mean values of the cross-power spectrum within defined $k$-bins),

\begin{equation}
[{\bfF}^{-1}]_{ij} =   \frac{1}{2}[{\bfW}^{-1}]_{ik} \left(W_{kl} + G_{kl} \right) [{\bfW}^{-1}]_{lj} = \frac{1}{2}\left ( [{\bfW}^{-1}]_{ij} + [{\bfW}^{-1}]_{ik} G_{kl}  [{\bfW}^{-1}]_{lj} \right ),
\label{eqn:fullFishCross}
\end{equation}
where
\begin{eqnarray}
\bfR_{i} &\equiv& \frac{\partial {\bfC^{\rm Ig}}}{\partial p_i};~~~~~~~~ W_{i j} = \frac{1}{2}{\rm Tr} \left [(\bfC^{\rm II})^{-1} \bfR_i (\bfC^{\rm gg})^{-1}\bfR_j^{\rm T} \right ];~~~ \\
G_{ij} &=&  \frac{1}{2}{\rm Tr} \Big[  (\bfC^{\rm II})^{-1} \bfR_i (\bfC^{\rm gg})^{-1} \bfC^{\rm gI} (\bfC^{\rm II})^{-1}\bfR_j  (\bfC^{\rm gg})^{-1} \bfC^{\rm gI} \Big].
\label{eqn:defns}
\end{eqnarray}
Here $\bfC^{\rm Ig}$ is the covariance matrix between our IM data $\bfI$ and our galaxy survey data $\bfg$ (which includes all pencil beams in the survey). Similarly, the covariance of the IM data with itself and the galaxy survey data with itself are denoted with $\bfC^{\rm II}$ and $\bfC^{\rm gg}$, respectively. In Appendix \ref{ap:covariances}, we derive all the elements of these matrices. In the limit of low noise, both terms in  Eq.~\ref{eqn:fullFishCross} contribute nearly equally \citep{2018PhRvD..98j3526V}, whereas in the more applicable case for this study of high noise the second term is more important.  

Let us specialize to the case where the parameters are the cross bandpowers we aim to detect; i.e. the $P_{\rm I g}(\bfk_\perp, k_\parallel)$ estimates that fall within some range of a wavevector. In Appendix~\ref{sec:crossestim}, we show that for this case in the limit where the \emph{JWST} fields are random in directions, the ensemble-averaged Fisher matrix error becomes 

\begin{equation}
\langle \delta P_{\rm I g}^2 \rangle =  \frac{1}{N_{\rm m}} \left ( \frac{V \,P_{\rm I}(\bfk_\perp, k_\parallel) \, \sigma_{\rm g}^{2}(k_\parallel)}{ d_\parallel^{2} \, N_{\rm g} \, \widetilde{W}_{\rm g}^{2}(\bfk_\perp)} + \frac{P_{\rm Ig}(\bfk_\perp, k_\parallel)^2}{2} \right ),
\label{eqn:fisherApproximation}
\end{equation}
where $N_{\rm m}$ is the number of modes in the bin around a given wavevector (counting only the Fourier modes where $k_\parallel >0$), $V$ is the IM survey volume, $P_{\rm I}$ is the IM power spectrum, $\sigma_{\rm g}^2(k_{\parallel}) = \langle {\rm Re}[g_a(k_{\parallel}) ]^2 \rangle $ is variance of galaxy overdensity in one pencil beam $a$ (computed in Appendix \ref{ap:covariances}), $d_\parallel$ is the line-on-sight length of the surveys, $N_{\rm g}$ is the number of pencil beams, $\widetilde{W}_{\rm g}$ is the Fourier transform of the pencil-beam window function, and $P_{\rm Ig}$ is the cross-power spectrum between galaxies and the IM survey.  For notational simplicity, Eq.~(\ref{eqn:fisherApproximation}) does not indicate the wavevectors used for $N_{\rm m}$ and $\delta P_{\rm I g}$. For the \emph{JWST} examples presented below, we demonstrate that this equation is an excellent approximation for the results given by the full Fisher formalism (Eq.~\ref{eqn:fullFishCross}).

In the limit that the galaxy shot noise dominates on the scale of the galaxy survey, we can further simplify Eq.~\ref{eqn:fisherApproximation}.

\begin{equation}
\langle \delta P_{\rm Ig}^2 \rangle =  \frac{1}{N_{\rm m}} \left( \frac{P_{\rm I}(\bfk_\perp, k_\parallel) }{2 f_{\rm cov} \ \bar{n}_g}   +  \frac{P_{\rm Ig}(\bfk_\perp, k_\parallel)^2}{2} \right),
\label{eqn:varshot2}
\end{equation}
where $f_{\rm cov}$ is the fraction of the IM field covered by JWST pointings (see Appendix \ref{sec:crossestim}).

There is one concerning aspect of the above equations.  The $P_{\rm Ig}(\bfk_\perp, k_\parallel)^2/(2N_{\rm m})$ terms in the previous two lined equations limits the S/N on a typical mode to be never be greater than $\sim 1$. This is because sample variance limits how well the cross-power spectrum can be constrained. However, we do not care about sample variance when asking how well cross correlations can be detected -- it is only the noise on the mode that matters, and so in principal the S/N of detecting correlation in a single mode can be detected can be arbitrarily large (but still one can only have an ${\cal O}(1)$-accurate estimate of the power spectrum that it was drawn from!).  In practice this distinction is  not very important, as most of the modes for the cases we consider are noise dominated. However, the above formula can be generalized to the case where one only cares about detecting the cross correlation and not constraining statically the value of the cross-power spectrum itself. In this case, one can show that the typical error to detect the cross correlations is approximately the same as Eq.~\ref{eqn:fisherApproximation} but dropping the $P_{\rm Ig}$ term and subtracting off the pure sample variance term in the auto (see Appendix~\ref{sec:signtonoisedetection} for the derivation). This results in a total S/N equal to $\sqrt{P_{\rm Ig}^2/\langle \delta P_{\rm Ig}^2 \rangle}$ on a typical mode of
\begin{equation}
\langle \delta P_{\rm Ig}^2 \rangle \approx \frac{V \, \left(P_{\rm I}(\bfk_\perp, k_\parallel) \, [\sigma_{\rm g}(k_\parallel)]^{2} - P_{\rm I}^{\rm SV}(\bfk_\perp, k_\parallel) \, [\sigma^{\rm SV}_{\rm g}(k_\parallel)]^{2}  \right)}{ N_{\rm m} d_\parallel^{2} \, N_{\rm g} \, \widetilde{W}_{\rm g}^{2}(\bfk_\perp)},
\label{eqn:noSampleApproximation}
\end{equation}
where now $P_{\rm I}^{\rm SV}$ and $\sigma^{\rm SV}_{\rm g}$ just include the signal and not the noise (where the noise is instrumental noise, uncorrelated shot noise, and noise due to foreground interloper lines).  In order to simplify to reach Eq.~\ref{eqn:noSampleApproximation}, we had to make an approximation that the covariance matrix in the galaxy pointings and the covariance matrix in the intensity mapping modes is diagonally dominated, which is most appropriate when each mode is still noise dominated (Appendix~\ref{sec:signtonoisedetection}).  These SV terms include correlated terms such as the part of each that traces the cosmic density field.  Noise due to foreground interloper galaxy lines (e.g., H$\alpha$ in the Ly$\alpha$ IM examples discussed below) must also be included (appearing in the noise component of $P_{\rm I}$). For the examples we show below, the total cross correlation S/N computed using Eq.~\ref{eqn:noSampleApproximation} is very similar to the S/N of the cross-power spectrum. However, there could be large differences in other cases of cross correlation that are less dominated by noise (both dector and galaxy shot noise).

\section{JWST Galaxy Observations}
\label{sec:JWST}
In this section, we describe our assumptions related to \emph{JWST} galaxy observations. While the formalism described above applies to any galaxy survey comprised of pencil beams, we focus on \emph{JWST} as an illustrative example. Throughout we consider a galaxy survey consisting of $N_{\rm g}$ pencil-beams, all within the IM survey volume being cross correlated. We focus on observations centered at $z = 7$ to show the utility of this cross-correlation technique during cosmic reionization. 

\subsection{JWST Sensitivity}
We assume that the \emph{JWST} galaxies are initially detected through Lyman-break selection in rapidly obtained snapshots with NIRCam (although this photometric survey can in detail be performed simultaneously with our NIRSpec observations). Following this initial detection, we assume that spectra are taken to estimate redshift values for each galaxy. We consider two separate cases for estimating redshifts: one based on detection of the Lyman-break and the other on Ly$\alpha$ line detection. We begin by describing the former. The redshifted Lyman-break for high-redshift galaxies occurs at $\lambda_{\rm obs} \approx (1+z)\times 1216~$ \AA ~ or $\approx 1 ~ {\rm micron}$ for $z\approx 7$ of interest. At this observed wavelength we require ${\rm S/N}= 3$ 
in a single spectral bin for an accurate redshift measurement, where we either use the instrumental resolution for the bin size or we combine nearby spectral pixels. ${\rm S/N} = 3$ means that roughly 95\% of the time the Lyman-break is located in the correct spectral pixel.  
These assumptions are intended to be a reasonable estimate for what is achievable with \emph{JWST}. We leave a more precise analysis of \emph{JWST}'s capabilities to future work.

We consider two different \emph{JWST} instrumental configurations for galaxy redshift measurements and utilize the \emph{JWST} exposure time calculator \citep{10.1117/12.2231768} to estimate their respective relevant limiting magnitudes. First, we consider using the NIRSPEC G140M/F100LP grating.  
This grating has an effective spectral resolution of ${\cal R} \equiv \nu/\Delta \nu \sim 700$, however for our analysis we take the combination of four of the actual spectral channels to be one spectral channel (only in the Lyman-break redshift measurements described here, not the Ly$\alpha$ case described below). This reduces our effective spectrum resolution, but increases the S/N by a factor of two. Thus, for this configuration we assume an effective spectral resolution of ${\cal R}=200$ and find that a 10-hour exposure of a $m_{\rm AB, 10hr}=26.8$ galaxy has a S/N of $3$ using the \emph{JWST} exposure time calculator. 
This sensitivity calculation assumes that the entire galaxy fits within each NIRSPEC shutter, which is likely since a shutter corresponds to a spatial extent of $1.1\times 2.6$ physical kpc$^2$ at $z=7$, somewhat larger than the half-light radius of each galaxy which HST observations find to be somewhat smaller than $1\,$physical~kpc, on average \citep{2013ApJ...777..155O}.  We assume that S/N scales with time due to photon counting statistics, such that our $1{-}{\sigma}$ galaxy magnitude goes as $m_{\rm AB, max} = m_{\rm AB, 10hr} + 2.5 \log{(\sqrt{t/10~{\rm hr}})}$. 
For our second instrument, we consider the NIRSpec MOS in PRISM mode, which we take to have ${\cal R} =30$ and a S/N of 3 at $m_{\rm AB, 10hr}=27.8$ after 10 hours of integration (the same time scaling as the previous instrument is assumed). This represents a more sensitive but lower spectral resolution method than the previous configuration. In all of the examples described below, we assume that each pencil-beam field is observed over an equal time $t_{\rm obs} = t_{\rm total}/N_{\rm g} - 20~ {\rm minutes}$.  Here we assume 20 minutes is spent for each pointing to slew the telescope and perform instrumental overheads. 
Slewing the telescope ${\sim}1$ degree (which is a typical mean separation of pencil beams in our examples below), takes approximately ${\sim}$10 minutes.\footnote{https://jwst-docs.stsci.edu/jppom/visit-overheads-timing-model/slew-times}
 Finding a guide star and performing onboard script system
compilation, exposure overhead, and visit cleanup takes an additional ${\sim}10$ minutes.\footnote{https://jwst-docs.stsci.edu/jwst-general-support/jwst-observing-overheads-and-time-accounting-overview/jwst-instrument-overheads}
We note that this accounting of time is meant to be approximate, and we defer a more precise estimate to future works.

In addition to redshift measurements via the Lyman break just described, we also consider examples where redshifts are obtained through detection of Ly$\alpha$ lines. In these cases,  we assume the high-resolution grating configuration of \emph{JWST} (${\cal R}=700$). A resolution of ${\cal R}=700$ corresponds to a velocity width of $430~{\rm km/s}$, which is likely broader then the typical (transmitted red side) Ly$\alpha$ line. Additionally, the bright Ly$\alpha$ lines we consider have much higher flux than the continuum in a single spectral element. 
In contrast to the Lyman-break case described above where many spectral pixels constrain the break, we assume that ${\rm S/N} = 5$ is required for redshift detection through the Ly$\alpha$ line since it likely falls in a single pixel. The limiting Ly$\alpha$ luminosity that can be detected is given by

\begin{equation}
    L_{\rm Ly\alpha, min} = 4\pi D_{\rm L}^2 f_{\rm AB} \Delta \nu_{\rm em} (1+z)^{-1} 10^{-m_{\rm AB, max}/2.5},
\end{equation}
where $D_{\rm L}$ is the cosmological luminosity distance, $f_{\rm AB} = 3.631 \times 10^{-20} ~{\rm erg~s^{-1}~Hz^{-1}~cm^{-2}}$, 
and $\Delta \nu_{\rm em}$ is the spectral width of the (${\cal R}=700$) frequency bin in the rest frame of the galaxy. As in the Lyman-break case, the limiting AB magnitude is given by $m_{\rm AB, max} = m_{\rm AB, 10hr} + 2.5 \log{(\sqrt{t/10~{\rm hr}})}$, but with $m_{\rm AB, 10hr}=25.5$ due to the higher S/N requirement and not grouping adjacent spectral pixels. As described below, we determine the number density of galaxies above the detection thresholds with observed Ly$\alpha$ luminosity functions.

\subsection{Galaxy Power Spectrum}
Once we specify $N_{\rm g}$ and $t_{\rm total}$ for a hypothetical survey, the galaxy sensitivity assumptions above provide us with the limiting observable magnitude, $m_{\rm AB, max}$.  This limiting magnitude is then used to determine the power spectrum of the observed galaxies, $P_{\rm g}(\bfk)$. We assume the power spectrum takes the standard clustered plus shot noise form: 

\begin{equation}
P_{\rm g}(\bfk) = \bar{b}_{\rm g}^2 P_{\rm m}(\bfk, z) + \frac{1}{\bar{n}_{\rm g}},   
\end{equation}
where $P_{\rm m}$ is the matter power spectrum, $\bar{b}_{\rm g}$ is the mean linear bias of the galaxies, and $\bar{n}_{\rm g}$ is the galaxies' comoving number density.  Because our calculations are in the low-number density, large-scale limit, this form for the power spectrum is likely a good approximation.  We use numerical values for $P_{\rm m}$ from the publicly available code CAMB.\footnote{https://camb.info/}  In the Lyman-break case, we obtain $\bar{n}_{\rm g}$ by integrating the galaxy UV luminosity function from \cite{2021AJ....162...47B} above $m_{\rm AB, min}$. For Ly$\alpha$, we integrate the Ly$\alpha$ luminosity function from \cite{2018ApJ...867...46I} above $L_{\rm Ly\alpha, min}$. We  estimate $\bar{b}_{\rm g}$ by halo abundance matching. This is accomplished by associating our observed number density with a minimum halo mass, $M_{\rm min}$, via

\begin{equation}
\bar{n}_{\rm g} = \int_{M_{\rm min}}^\infty  dM \frac{dn}{dM} \epsilon_{\rm duty},
\end{equation}
where  $\epsilon_{\rm duty}$ is the galaxy duty cycle and $\frac{dn}{dM}$ is the halo mass function, for which we use the Sheth-Tormen fit to N-body simulations \citep{2001MNRAS.323....1S}. Once this minimum mass is determined an associated mean bias is computed with 

\begin{equation}
\bar{b}_{\rm g} = \left ( \int_{M_{\rm min}}^\infty dM b_{\rm ST}(M) \frac{dn}{dM} \right )/ \left (\int_{M_{\rm min}}^\infty  dM \frac{dn}{dM} \right ),
\end{equation}
where $b_{ST}$ is the Sheth-Torman linear bias as a function of halo mass \citep{2001MNRAS.323....1S}. For the Lyman-break galaxies we assume $\epsilon_{\rm duty}=1$, and for the Ly$\alpha$ we assume $\epsilon_{\rm duty}=0.05$. The latter yields a mean bias similar to observations of Ly$\alpha$ emitters at $z=6.6$ \citep{2018PASJ...70S..13O}.
Note that even though the duty cycle is not mass dependent, it changes the bias because it sets $M_{\rm min}$ and higher $M_{\rm min}$ then results in higher bias.

\begin{figure}
     \centering
     \includegraphics[width= 8cm]{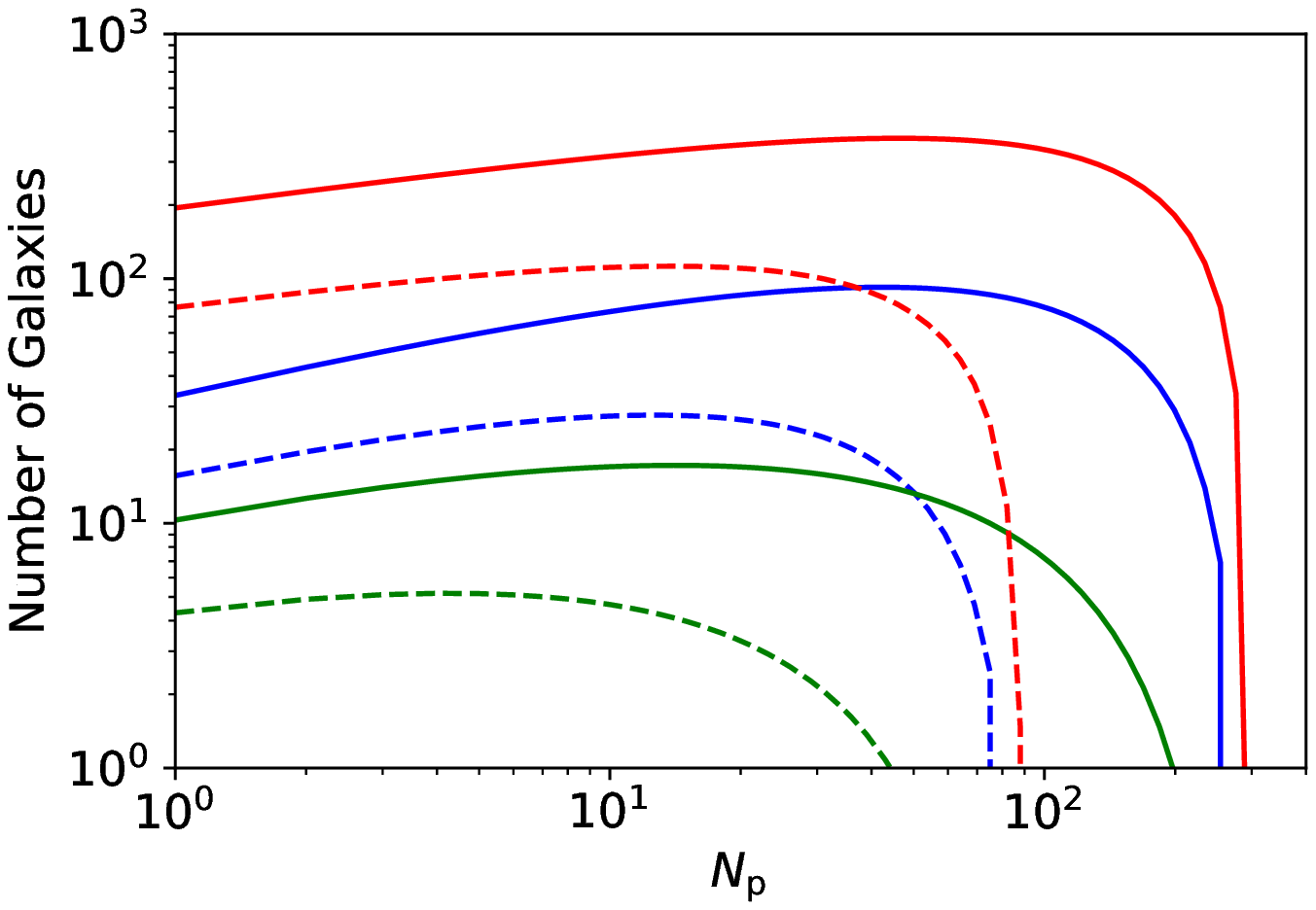}
     \includegraphics[width= 8cm]{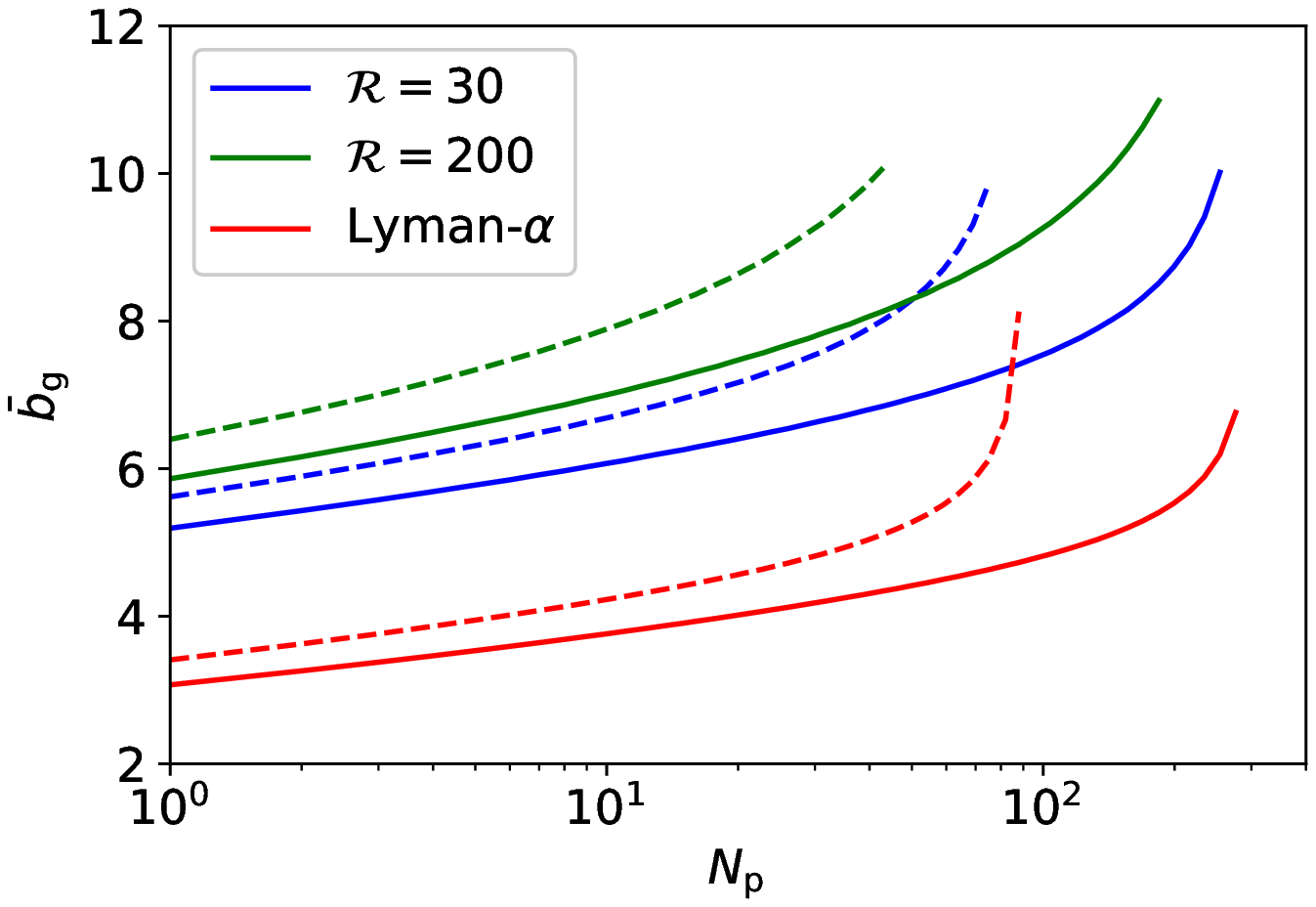}
     \caption{\label{fig:bias_and_density} The number of galaxies with redshifts detected by \emph{JWST} (left) and the corresponding mean bias (right), as a function of \emph{JWST} pointings, $N_{\rm p}$, observed for 100 hours (solid curves) and 30 hours (dashed curves) of total time. This includes 20 minutes of slew time and instrumental overhead for each field. We show results for our three methods of redshift detection: low-resolution Lyman-break (${\cal R} = 30$), high-resolution Lyman-break (${\cal R} = 200 $), and Ly$\alpha$ line detection.  
     }
 \end{figure}

In Figure \ref{fig:bias_and_density}, we plot the number of galaxies with measured redshifts and the corresponding $\bar{b}_{\rm g}$ as a function of the number of \emph{JWST} fields. We find that out of the three considered methods for obtaining galaxy redshifts, Ly$\alpha$ detection is the most sensitive, followed by the low-resolution Lyman-break technique, and finally the high-resolution Lyman-break. For 100 total hours of \emph{JWST} time with the Ly$\alpha$ technique, we find ${\sim}100$ pointings appears to maximize the number of detected galaxies. We point out that this is very similar to the same area on the sky covered by the COSMOS-Webb survey \citep{2022arXiv221107865C}, which will cover ${\sim} (90 Mpc)^2$ or $(0.6 {\rm deg})^2$. However, in our examples presented below this same total area would be spread out over the larger footprints of the respective IM surveys (e.g., $(6.3 {\rm deg})^2$ in the case of \emph{SPHEREx}-like experiment).

In the cross-correlation sensitivity calculations performed below, we assume a square \emph{JWST} FOV is covering a distance $L=8 ~{\rm Mpc}$ on each side (comoving).  This  corresponds to the ${\sim}$9 arcmin$^2$ FOV of NIRSpec.
Thus, the window function appearing above and in Appendix~\ref{ap:covariances} is given by $\tilde{W}(k_\perp)_g = \sinc{(k_x L/2)} \sinc{(k_y L/2)}$, where $k_x$ and $k_y$
are components of the wavevector perpendicular to the line-of-sight. 
The comoving length covered by each pencil beam along the line-of-sight, $d_{\parallel}$, and comoving distance associated with one frequency channel sets the available wavevector modes along the line-of-sight, $k_{\parallel, i}$.  Generally, these range from  $k_{\parallel} = 0$ to $k_{\rm max, \parallel} = \pi/(\Delta d_{\parallel}$) with a resolution of $\Delta k_{\rm \parallel} = 2 \pi/d_{\parallel}$.\footnote{However, as discussed later, in the case of 21cm observations, many of these wavevectors cannot be used due to the ``wedge'' in $k$-space contaminated by foreground removal.} Note that the distance corresponding to a frequency channel is given by  $\Delta d_{\parallel} \approx  \frac{\lambda_{\rm_{obs}}(1+z)}{H(z)}\Delta \nu$,
where $H(z)$ is the Hubble parameter and $\lambda_{\rm obs}$ is the observed wavelength.

\section{JWST Pencil Beam-Galaxy Line-Intensity Cross Correlations}
\label{sec:sensitivity}

\subsection{IM-Galaxy Cross-Power and IM Power Spectra}
In order to demonstrate the utility of the cross-correlation technique described above, we explore combining \emph{JWST} pencil-beam galaxy surveys with a sample variance-limited (SVL) survey of galaxy-line emission and Ly$\alpha$ surveys with an instrument similar to \emph{SPHEREx}.
We utilize the formalism in Section \ref{sec:formalism} to estimate the sensitivity of the IM-galaxy cross-power spectrum. We assume that this power is given by

\begin{equation}
P_{\rm  Ig}(k) = b_{\rm I} \bar{S}_{\rm I} b_{\rm g} P_{\rm m}(k) + P_{\rm cross-shot},
\end{equation}
where $b_{\rm I}$ is the mean bias of the IM survey, $\bar{S}_{\rm I}$ is the mean intensity of the IM signal, and $P_{\rm cross-shot}$ is the cross-shot noise power spectra due to the overlapping shot noise from the two surveys. It is zero in the limit that the IM surveys owes to much smaller galaxies than \emph{JWST} can observe. The first term is the clustering term and thus is proportional to the matter power spectrum, $P_{\rm m}$. For the SVL and Ly$\alpha$ IM surveys, we assume that the flux from each galaxy is proportional to its host dark matter halo's mass and that there is signal from halos above a minimum mass $M_{\rm min, I} = 1.5\times 10^9~M_\odot$, following \cite{2018ApJ...863L...6V}. Note that for the IM surveys we consider this minimum mass is generally smaller than the minimum detectable halo mass detected in the \emph{JWST} survey. With these assumptions the luminosity-weighted bias is given by 

\begin{equation}
\bar{b}_{\rm I} = \left ( \int_{M_{\rm min,I}}^\infty dM b_{\rm ST}(M)M \frac{dn}{dM} \right )/ \left (\int_{M_{\rm min, I}}^\infty  dM M \frac{dn}{dM} \right ),
\end{equation}
and the mean signal by 

\begin{equation}
\label{avgsig}
\bar{S}_{\rm I}=\int_{M_{\rm min, I}}^{\infty} dM\frac{L(M)}{4\pi D_L^2} \epsilon_{\rm duty, I} \frac{dn}{dM}\tilde{y}D_A^2,
\end{equation}  
where $D_{\rm A}$ is the angular diameter distance, $D_{\rm L}$ is the luminosity distance, $\epsilon_{\rm duty, I}$ is the duty cycle of the galaxies contributing to the IM signal, and $\tilde{y}$ is the derivative of the comoving distance with respect to the observed frequency ($\bar{S}_{\rm I}$ then has units of spectral flux density per solid angle). We assume $L(M) \propto M$, but note that the constant of proportionality does not impact the S/N in the SVL case because it appears in both the signal and noise. Note that $L(M)\propto 1/\epsilon_{\rm duty, I}$, such that the mean IM signal does not depend on the duty cycle.
For the cross-shot power in the SVL case, we both make the conservative assumption that there is no cross-shot power as well as the maximal case where the exact same galaxies detected with \emph{JWST} also source the intensity maps yielding 

\begin{equation}
\label{eqn:crossShot}
P_{\rm cross-shot}=\frac{1}{ \bar{n}_g}  \frac{\epsilon_{\rm duty}}{\epsilon_{\rm duty, I}} \bar{S}_{\rm I}.
\end{equation}  
 Here the ratio of the duty cycles accounts for the for the fact that either some of the intensity mapping galaxies do not contribute to the galaxy detections (if ${\epsilon_{\rm duty, I}} > \epsilon_{\rm duty}$) or some of the detected galaxies do not contribute to the intensity map (if ${\epsilon_{\rm duty, I}} < \epsilon_{\rm duty}$).
We also require the IM auto-correlation power spectrum for our sensitivity calculations, which is given by 

\begin{equation}
P_{\rm I} = b_{\rm I}^2 \bar{S}_{\rm I}^2 P_{\rm m} + P_{\rm I, shot} + P_{\rm N},
\end{equation}
where 

\begin{equation}
P_{\rm I, shot} = \int_{M_{\rm min,IM}}^{\infty} dM \left ( \frac{L(M)}{4\pi D_L^2}  \tilde{y}D_A^2 \right )^2 \frac{dn}{dM} {\epsilon_{\rm duty, I}},
\end{equation}
and $P_{\rm N}$ is the power due to detector noise \citep{2010JCAP...11..016V}. The duty cycle of the line emitting galaxies that contribute to the IM signal is given by $\epsilon_{\rm duty, I}$, which we assign values of 1 and 0.1 for our sample variance-limited and Ly$\alpha$ IM survey examples, respectively. We note that the Ly$\alpha$ IM duty cycle is twice as large as the duty cycle used to estimate the clustering bias of the \emph{JWST}-detected Ly$\alpha$ emitting galaxies. However, this is reasonable since the duty cycle is for different populations of galaxies; the Ly$\alpha$ IM signal mostly comes from  faint and abundant galaxies which we do not detect directly, as opposed to the brighter galaxies observed directly with \emph{JWST}.
We note that throughout we have ignored the impact of the redshift-space distortions on our power spectra. Given that all of the relevant biases are typically ${\gtrsim} 4$, we do not expect this to strongly change our results. We also note that this is a conservative choice in the sense that redshift-space distortions would increase the clustering signal relative to the noise power, increasing the S/N in the examples below.

\subsection{Sample Variance Limited Examples}
In this subsection, we present a series of results for an IM survey containing negligible detector noise, which we refer to as the SVL IM survey. These examples are intended to show the behavior of the sensitivity with respect to number and configuration of pencil beams and to demonstrate the accuracy of the approximation given by Eq.~\ref{eqn:fisherApproximation}. They also estimate that maximum S/N in this most idealized limit of no noise in the IM survey.  More realistic cases with detector noise are explored later. 

We consider the noise-free IM survey to be centered at $z=7$ and to span a square FOV with comoving area of $({\rm 300\; Mpc})^2$ and a depth of $d_\parallel = 150 \;{\rm Mpc}$ (corresponding to $\Delta z \approx 0.5$). We assume that the angular resolution matches the size of our 8 Mpc$\times$8 Mpc \emph{JWST} pencil beam FOV and the spectral resolution matches \emph{JWST} for which redshift detection mode is being explored. In each transverse dimension, the $k$-space resolution is $\Delta k_{\rm x} = 2\pi/L_{\rm x}$, where $L_{\rm x}$ and $\Delta k_{\rm x}$ are the size and spatial resolution of the IM survey in that dimension, and the angular wavemodes span $\pm \pi/L$ (where $L=8$ Mpc).

\begin{figure}
     \centering
     \includegraphics[width= 8cm]{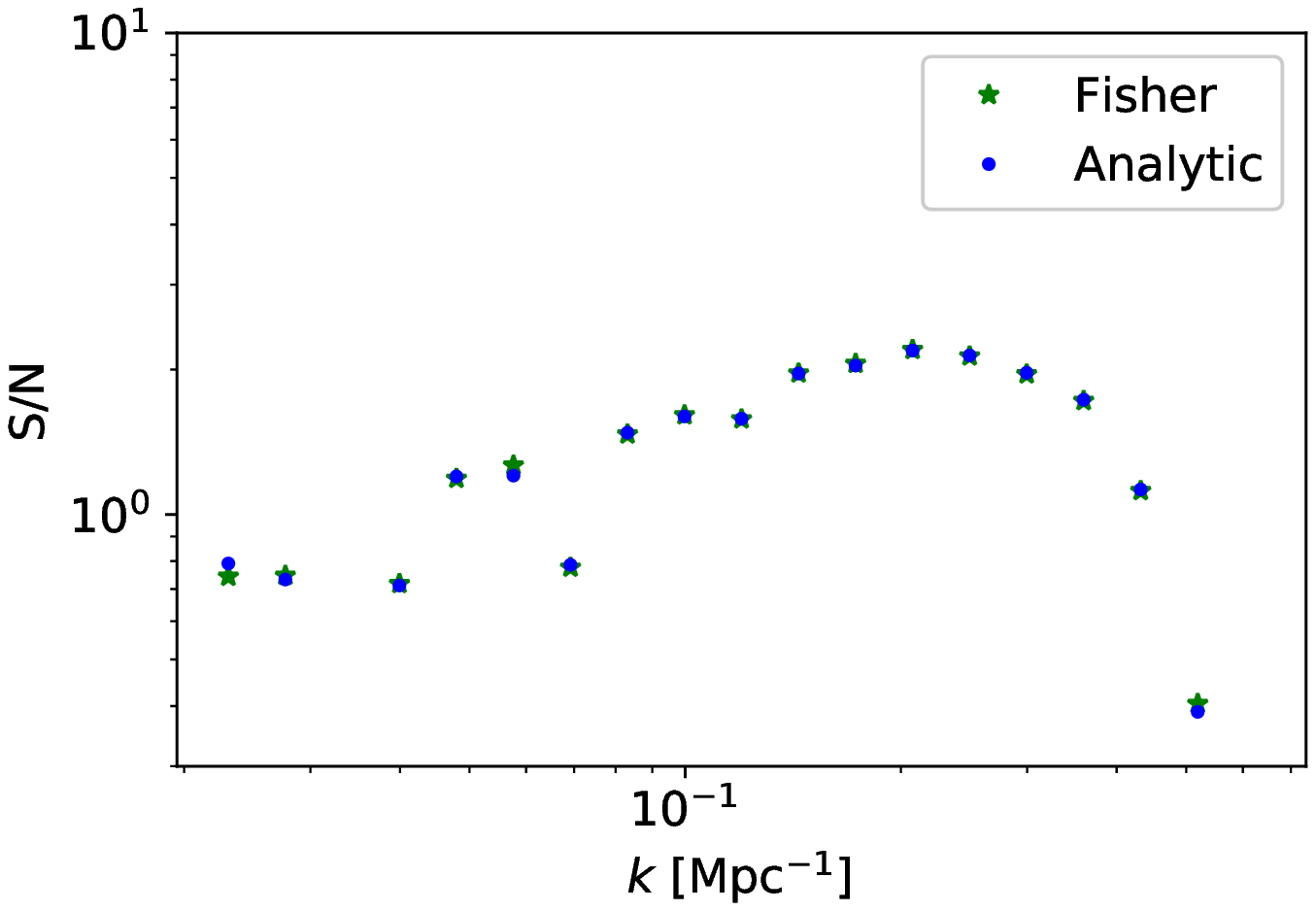}
     \includegraphics[width= 8cm]{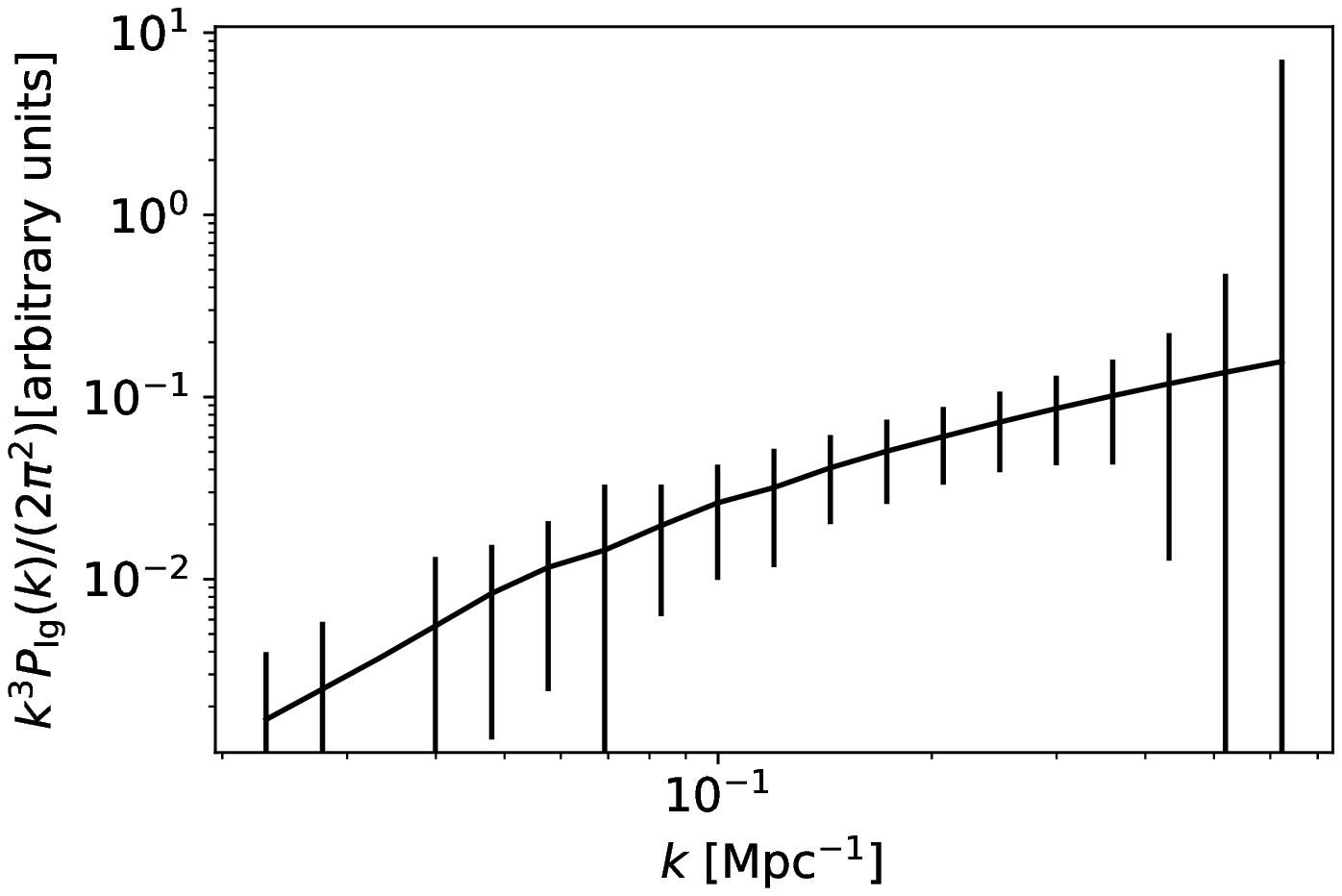}
     \caption{\label{cvl_check1} The illustrative case of correlating \emph{JWST} with a SVL IM survey (i.e.~one where the instrumental noise is zero).  \emph{Left Panel}: The S/N of the cross-power spectrum of IM survey and $N_{\rm g}=25$ \emph{JWST} pencil beans with 100 hours of total integration time as a function of $k$. We include both the S/N computed with our full Fisher matrix calculation (Eqn. \ref{eqn:fullFishCross}) and the analytic approximation in Eqn. \ref{eqn:fisherApproximation}. For most of the points plotted, they agree better than ${\sim}1$ percent. \emph{Right Panel}: The value of the cross-power spectrum and the associated error bars (computed with the Fisher matrix formalism). The units are arbitrary since the S/N does not depend on the overall normalization of the cross-power spectrum when there is zero detector noise.  The non-uniform size of the error bars with wavenumber owes to variations in the number of modes that fall into a band power bin.
     }
\end{figure}

We begin by examining cross correlation between the SVL survey and 25 randomly positioned \emph{JWST} pencil beams split across 100 hours of observing time (including the 20 minutes spent switching between each FOV). Redshifts are assumed to be measured with the ${\cal R}=200$ Lyman-break mode of \emph{JWST}. In Figure \ref{cvl_check1}, we show the S/N of the binned cross-power spectrum as a function of wavenumber for both the full Fisher calculation given by Eq.~\ref{eqn:fullFishCross} and the approximation given by Eq.~\ref{eqn:fisherApproximation}, where for each bandpower (S/N)$^2 = P_{\rm Ig}^2/\langle \delta P_{\rm Ig}^2  \rangle$. 
We find that the approximation in Eq.~\ref{eqn:fisherApproximation} matches the Fisher matrix calculation better than ${\sim}1$-percent for most of the $k$-bins in Figure \ref{cvl_check1}, though we note the discrepancy is ${\sim}10$ percent in the lowest $k$-bin. We find similar agreement when adding noise for our other examples.

\begin{figure}
     \centering
     \includegraphics[width= 8cm]{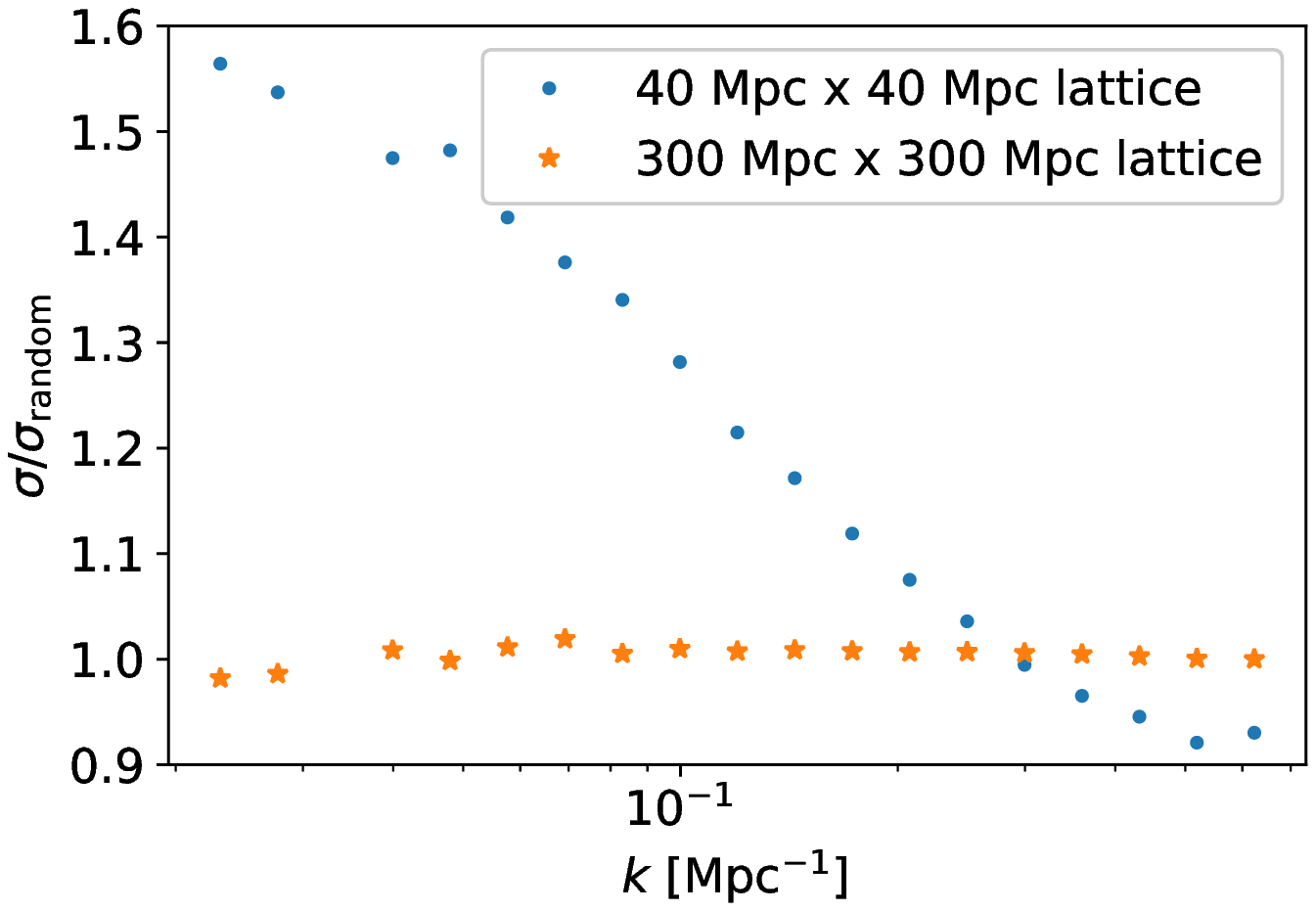}
     \includegraphics[width= 6cm]{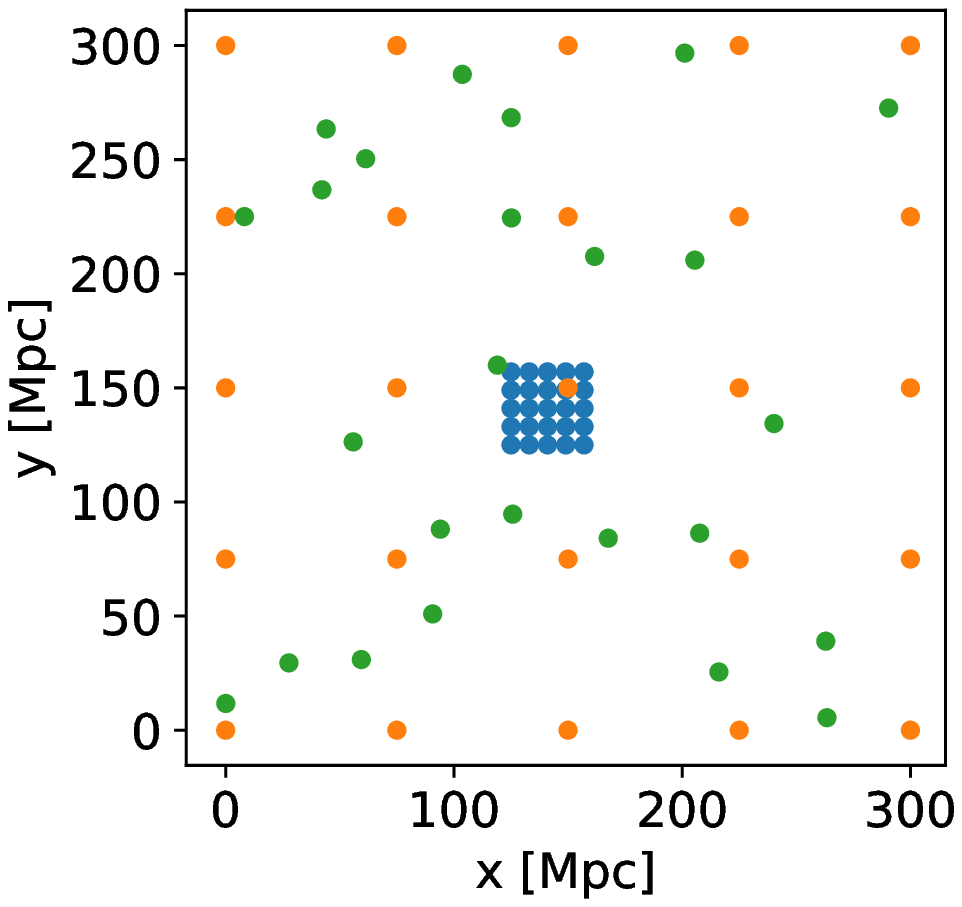}
     \caption{ Impact of \emph{JWST} field configuration on the error of the galaxy-IM cross-power spectrum. As in Figure \ref{cvl_check1}, we consider 25 \emph{JWST} fields distributed within our SVL IM survey with 100 hours to total \emph{JWST} time. We consider three configurations: a `random' distribution, a lattice spread evenly across the IM survey, and a `tight' lattice with the fields adjacent to one another and only spanning $(40 {\rm Mpc})^2$ of the $(300 {\rm Mpc})^2$ IM survey (see diagram in right panel; note that the size of points in this diagram are not meant to match the 8 Mpc ${\times}$ 8 Mpc \emph{JWST} FOV). We plot the ratio of the tight and grid errors with the random case (left panel).
     All calculations are done with the full Fisher matrix formalism in Eq. \ref{eqn:fullFishCross}. 
     }
     \label{grid_check}
\end{figure}

Next, we compare the impact of the positioning of the \emph{JWST} pencil beams within the IM field. In Figure \ref{grid_check}, we show the S/N of the cross-power spectrum for the extreme cases of a lattice versus a random distribution where both are set to cover the IM survey area spanning $300 ~{\rm Mpc} \times 300 ~{\rm Mpc} $. See the right panel for visualization of these configurations.  We find that the two cases lead to similar sensitivities, with the lattices only slightly improving the S/N at the lowest wavenumbers.  The reason for the small differences is the projection of 3D modes onto 2D; if we were instead interested in 2D modes the lattice would certain favor some more than others. We also compare the case of a compact lattice. In this example, the \emph{JWST} fields are tightly packed with a spacing of 8 Mpc such that it spans spanning $40 ~{\rm Mpc} \times 40 ~{\rm Mpc}$  of the $300 ~{\rm Mpc} \times 300 ~{\rm Mpc} $ field. As in the previous example, we assume 100 hours with \emph{JWST} in the high-resolution Lyman-break mode described above. We find  that as expected the compact configurations improves sensitivity on small scales by ${\sim} 3$\% and decreases sensitivity on large scales, with this difference reaching ${\sim} 50$\% at the smallest wavenumber we consider.

\begin{figure}
     \centering
     \includegraphics[width= 8cm]{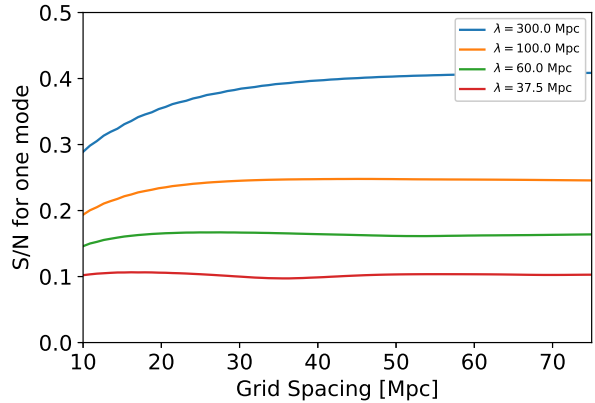}
     \caption{\label{cvl_one_mode}S/N of the cross-power spectrum when only including one SVL IM Fourier mode and for different grid spacings covering a subset of the IM field. The mode is assumed to be perpendicular to the line of sight and different wavelengths given by $\lambda = 2\pi/k_\perp$ are explored, although the sensitivity for inclined modes are similar as long as the parallel wavenumber is less than $k_\perp$. The pencil-beam field survey consists of 25 fields arranged in a grid with uniform spacing that varies from compact to covering the entire IM survey (assuming 100 hours of \emph{JWST} time and the high-resolution Lyman-break mode for redshift measurement). The sensitivity computation is completed with the full Fisher formula given in Eq. \ref{eqn:fullFishCross}. A ${\rm S/N}=1$ corresponds to the best which is possible due to sample variance. Large-wavelength modes are best measured by wide grid spacing, as one would expect. }
 \end{figure}

We also wish to examine how the configuration of \emph{JWST} fields impacts the sensitivity of individual modes. In Figure \ref{cvl_one_mode}, we show the S/N of the cross-power spectrum for one mode perpendicular to the line-of-sight with varying wavelength $\lambda = 2\pi/k_\perp$ and grid spacing.
As in the previous examples, we assume 25 \emph{JWST} fields over 100 hours, using the high-resolution Lyman-break technique to measure galaxy redshifts.  The square grid of pointings cover a subset of the entire field until a spacing of $75~$Mpc.  
As expected, we find that smaller spacing modestly increases the sensitivity to small-scale (high-$\bfk$) modes  of the power spectrum and reduces the sensitivity more significantly on large scales.

In our final test of the SVL IM survey, we explore the cross-power spectrum sensitivity as a function of correlating with different numbers of randomly positioned \emph{JWST} fields (but with fixed total observation time). Here define the total S/N as the square root of the sum of the S/N squared in all wavevector bins. We compute the S/N with the random-field approximation from Eq.~\ref{eqn:fisherApproximation}, as justified by the previous results in this subsection. Operationally, we break $k$-space up into a number of 2-dimensional bins spanning the magnitudes of the wavevector, $|\bfk|$, and the angles offset from the line-of-sight, $\theta$ (where $\tan (\theta) \equiv k_\perp / k_\parallel$). We then estimate the band cross-power error in each $k$-bin by taking the inverse variance weighted average of the error implied by Eq.~\ref{eqn:fisherApproximation} averaged over our $|\bfk|$ and $\theta$ bins. 

\begin{figure}
     \centering
     \includegraphics[width= 8cm]{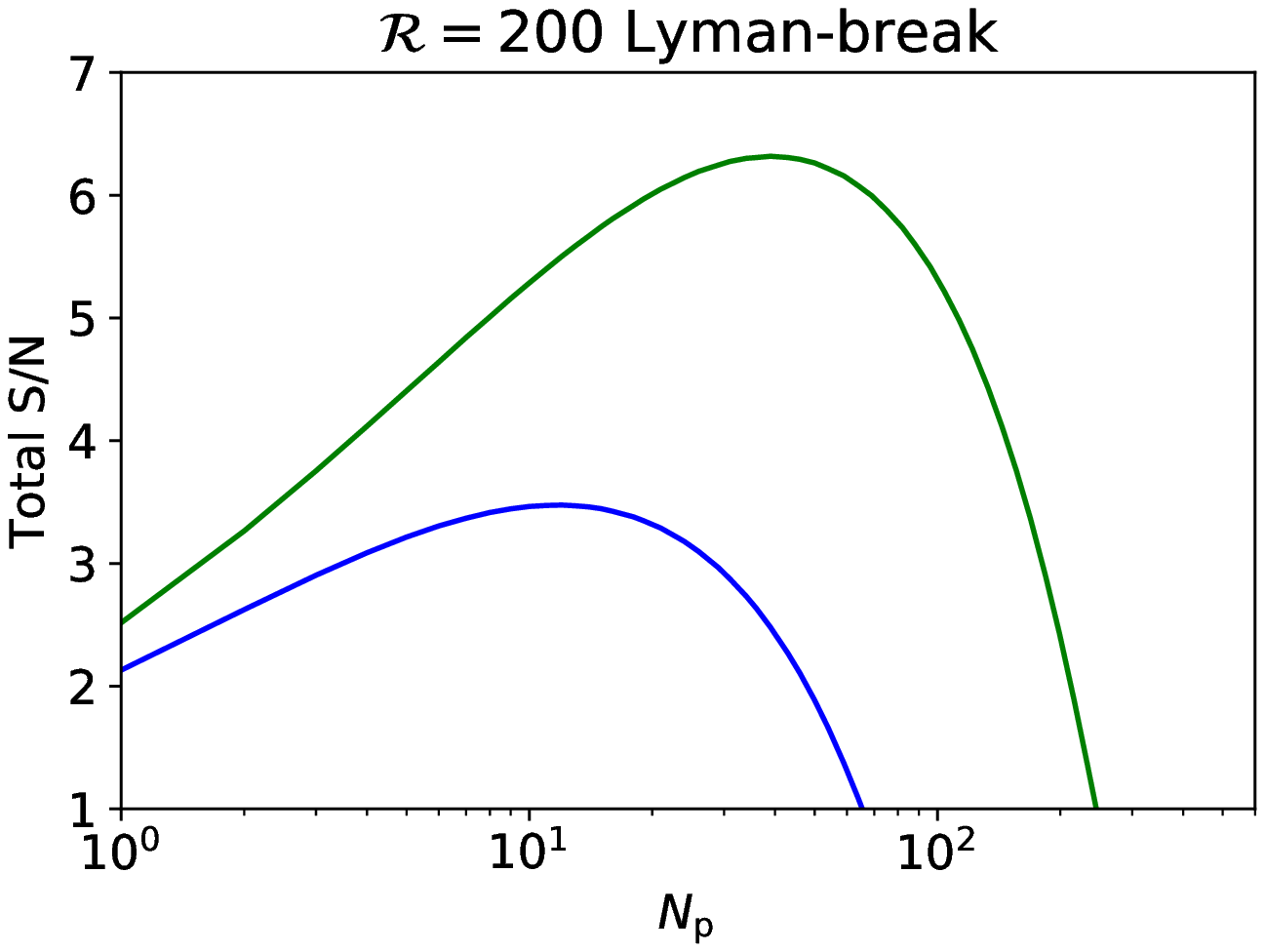}
     \includegraphics[width= 8cm]{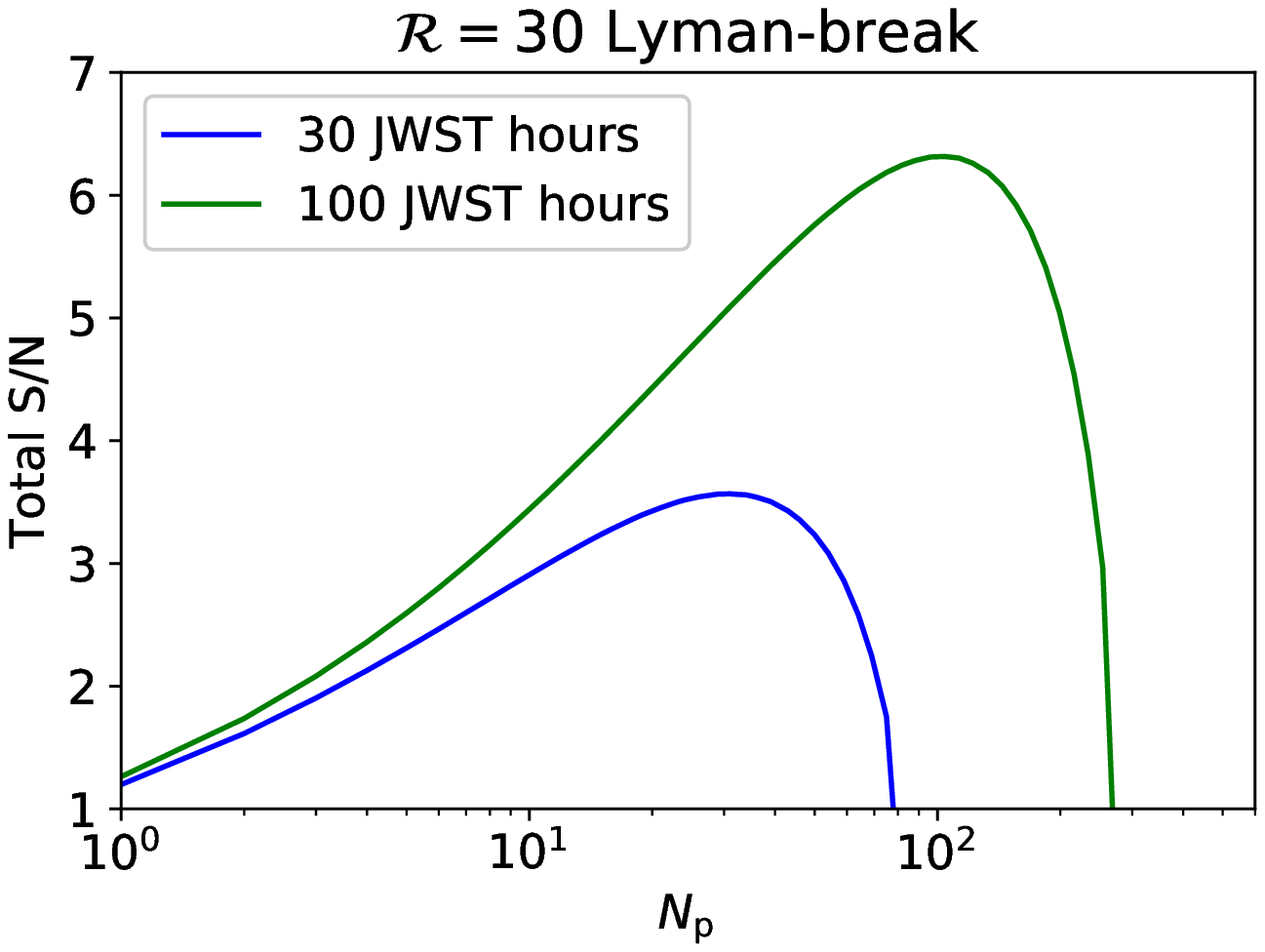}
     \includegraphics[width= 8cm]{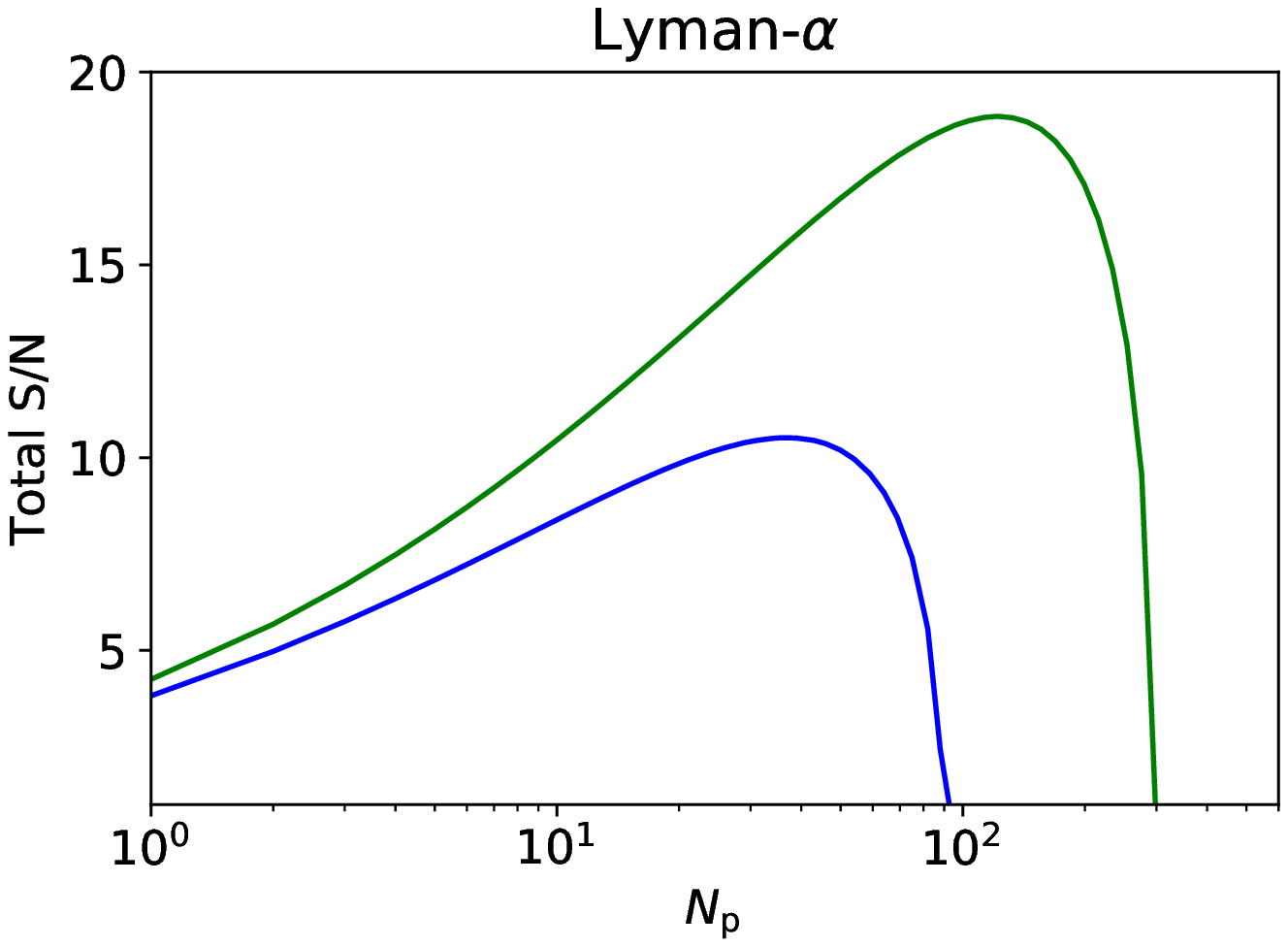}
     \caption{\label{cvl_check2} The total S/N (defined as the square root of the S/N in our $k$-bins added in quadrature) on the cross-power spectrum for galaxies detected with \emph{JWST} and our SVL IM survey (computed with Eq.~\ref{eqn:fisherApproximation}). We show the total S/N for 30 (blue curves) and 100 hours (green curves) of total \emph{JWST} observation times varying the number of randomly positioned pencil beams, $N_{\rm p}$. Results are included for each of the three ways of measuring galaxy redshifts described in Section \ref{sec:JWST}: ${\cal R} =200$ Lyman-break, ${\cal R} =30$ Lyman-break, and Ly$\alpha$. The S/N is maximized for $N_{\rm p}$ similar to that where the most galaxy redshifts are detected.
     }
\end{figure}

In Figure \ref{cvl_check2}, we present the total S/N on statistical measurements of the cross-power spectrum (solid curves). As described above these are well approximated with Eq.~\ref{eqn:fisherApproximation}. 
The blue curves assume 30 total hours of \emph{JWST} observations and the green $100$ hours, and the different panels consider the three galaxy redshift identification methods.  In all of the cases, we find that the S/N is maximized roughly when the number of galaxies are maximized (see Figure~\ref{fig:bias_and_density}). 
We find that obtaining galaxy redshifts with the Ly$\alpha$ line has the highest S/N, followed by the low-resolution Lyman-break technique and then the high-resolution Lyman-break technique. 
In 100 hours of integration time, S/N in the Ly$\alpha$ case of ${\gtrsim}10$ is achieved in the cross-power spectrum. These S/N bound what is achievable in a realistic case with IM noise.

We have not included the correlations between the shot noise of the two surveys (the `cross-shot power') in any of the calculations shown in Figure \ref{cvl_check2}. We note that including cross-shot noise power only makes a modest  difference. For example, in the Ly$\alpha$ case the peak S/N is increased by a factor of ${\sim}1.1$ when we assume the maximal case where the shot noises of the two surveys are perfectly correlated. In the Ly$\alpha$ IM survey presented below there is a more significant impact, due to the large ratio of $\frac{\epsilon_{\rm duty}}{\epsilon_{\rm duty, I}}$ considered.

\subsection{Ly$\alpha$ IM with SPHEREx} 
In this subsection, we explore cross correlation of \emph{JWST} pencil beams with a Ly$\alpha$ IM survey measured by an instrument similar to \emph{SPHEREx}. We focus on a case where the IM survey is centered at $z=7$ and covers an area of $(1000~{\rm Mpc})^2$ on the sky (corresponding to the \emph{SPHEREx} 40 sq.~deg. FOV), and a depth of $d_\parallel = 150 \,{\rm Mpc}$ (corresponding to $\Delta z \approx 0.5$). 
We follow \cite{2018ApJ...863L...6V} to estimate the mean signal of the IM power spectrum.

The Ly$\alpha$ luminosity of a galaxy is related to its star formation rate, $\dot{M}_*$, by

\begin{equation}
\label{luminosity}
L_{\rm gal}=2\times10^{42}\left(1-f_{\rm esc}\right)\frac{\dot{M}_*}{M_\odot~{\rm yr}^{-1}}~{\rm erg~s^{-1}},
\end{equation}
where $f_{\rm esc}$ is the fraction of ionizing photons that escape into the IGM. This equation assumes there is no dust absorption, such that every ionization results in $0.6$ Ly$\alpha$ photons. It also assumes a Salpeter initial mass function (IMF; \citealt{2003A&A...397..527S}) over a mass range of $1-100 M_\odot$ with metallicity $Z=0.04$; other empirically-motivated PopII IMFs result in factor of $\sim 2$ differences. We assume that a fraction $\epsilon_{\rm duty }=0.1$ of halos are forming stars at any specific time and that $\dot{M}_*$ is proportional to halo mass, with a normalization such that the global star formation rate density at $z=7$ is 
$\rho_\star=0.015~M_\odot~ {\rm yr^{-1}~Mpc^{-3}}$. This value is similar to $\rho_\star = 0.02~M_\odot~{\rm yr^{-1}~Mpc^{-3}}$ measured by \cite{2015ApJ...803...34B} at $z\approx 6.8$. 
We have assumed that there is no scattering of Ly$\alpha$ photons by the IGM. This would be a good assumption if reionization is nearly complete. For high cosmic neutral fractions of hydrogen scattering would suppress small-scale power \citep{2018ApJ...863L...6V}, but we do not expect strong suppression at the scales we are sensitive to at $z{\sim}7$.

We estimate the power of instrumental noise using Eq.~16 from 
\cite{2016MNRAS.463.3193C}.  To approximate the \emph{SPHEREx} specifications, we assume a telescope diameter of 20 cm, a zodiacal light background intensity of $\nu I_\nu=500~{\rm nW~m^{-2}~sr^{-1}}$, 
and an observational efficiency of detecting a photon accounting for losses in the instrument of $\epsilon = 0.5$. We assume an integration time of $10^6$ seconds. 
Additional noise in the cross-power comes from foreground interloper lines in the IM. In our example, this noise is expected to be dominated by foreground H$\alpha$ (appearing at $z\approx 0.5$ for our survey at $z\approx 7$). Following \cite{2014ApJ...786..111P} (see their Figure 13), we assume that this signal has power given by $P_{\rm H\alpha}=0.04\times({\rm{Mpc}^{-1}}/k)~{\rm{nW}~m^{-2}~sr^{-1}Mpc^3}$. This assumes that the sources brighter than $10^{-17}~{\rm{erg}~s^{-1}~cm^{-2}}$  have been identified and the their contributions removed from the IM map. This flux cut corresponds to an r-band AB magnitude of $m_r\approx26.5$, which will be observable over large areas with telescopes such as the Hyper Suprime-Cam \citep{2014ApJ...786..111P}.

We consider two different methods for galaxy redshift measurements of our \emph{JWST} pencil beams (discussed in detail above): the lower-resolution PRISM mode to detect the Lyman-break and the high-resolution mode to detect Ly$\alpha$ line emission. We assume the spectral resolution of \emph{SPHEREx} is ${\cal R}=40$, which does not have an impact on the ${\cal R}=30$ PRISM mode Lyman-break examples (in this case we assume the IM spectral pixels are combined to match the PRISM resolution), but sets the line-of-sight spatial resolution for cross-correlation between \emph{JWST} galaxies and the IM when redshifts are obtained from Ly$\alpha$ lines. 

\begin{figure}
     \centering
     \includegraphics[width= 7.8cm]{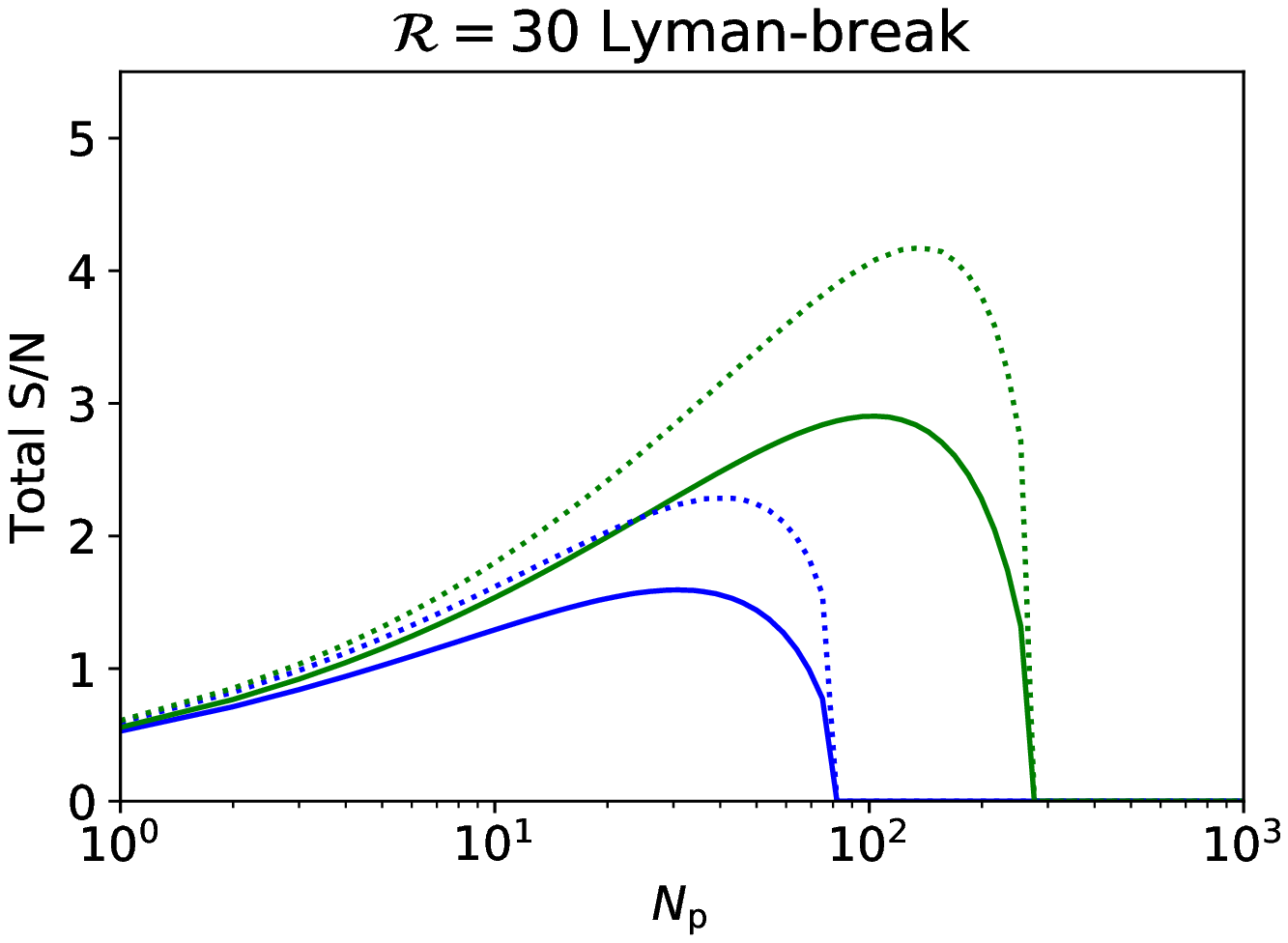}
     \includegraphics[width= 8cm]{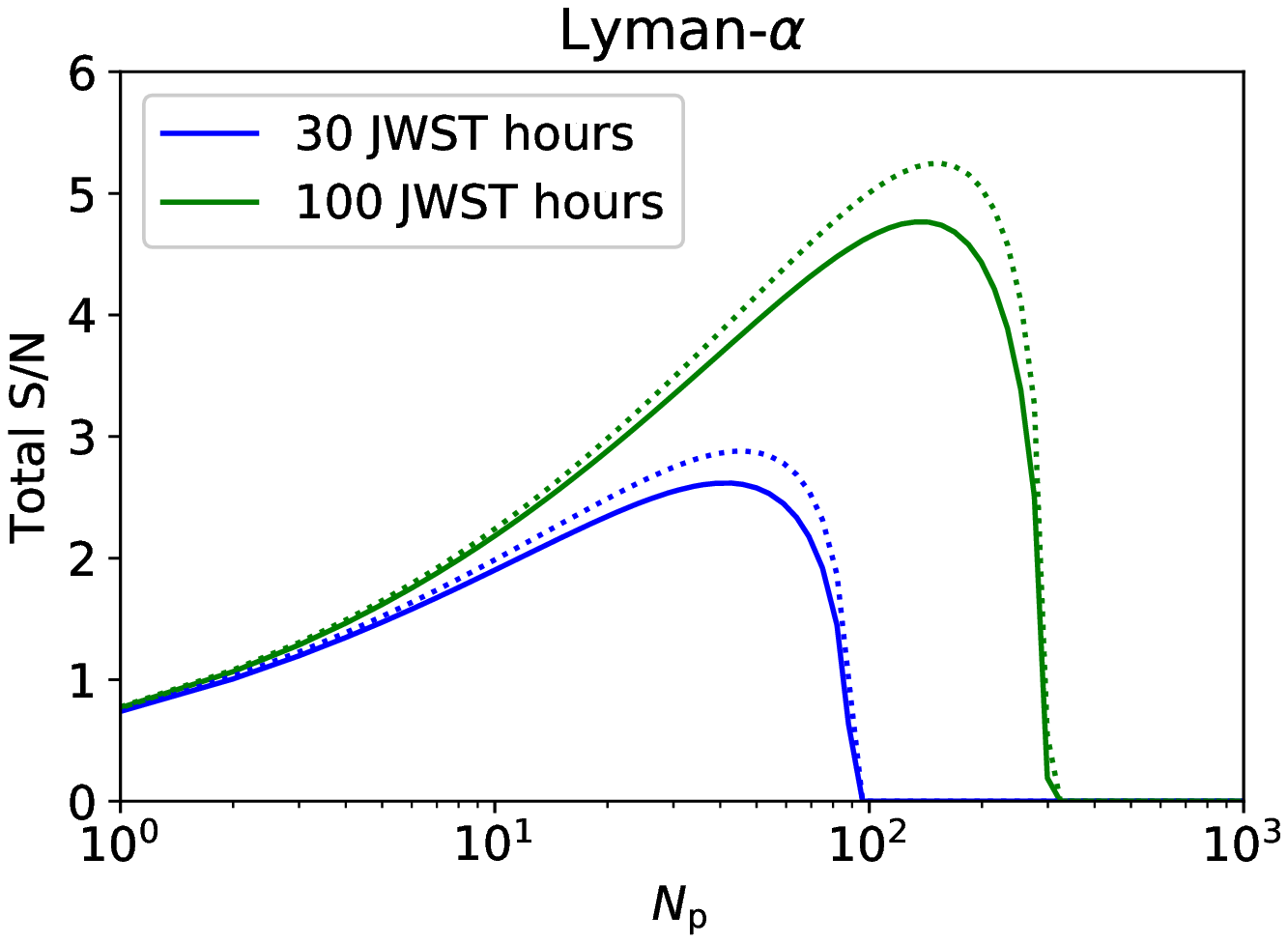}
     \includegraphics[width= 8cm]{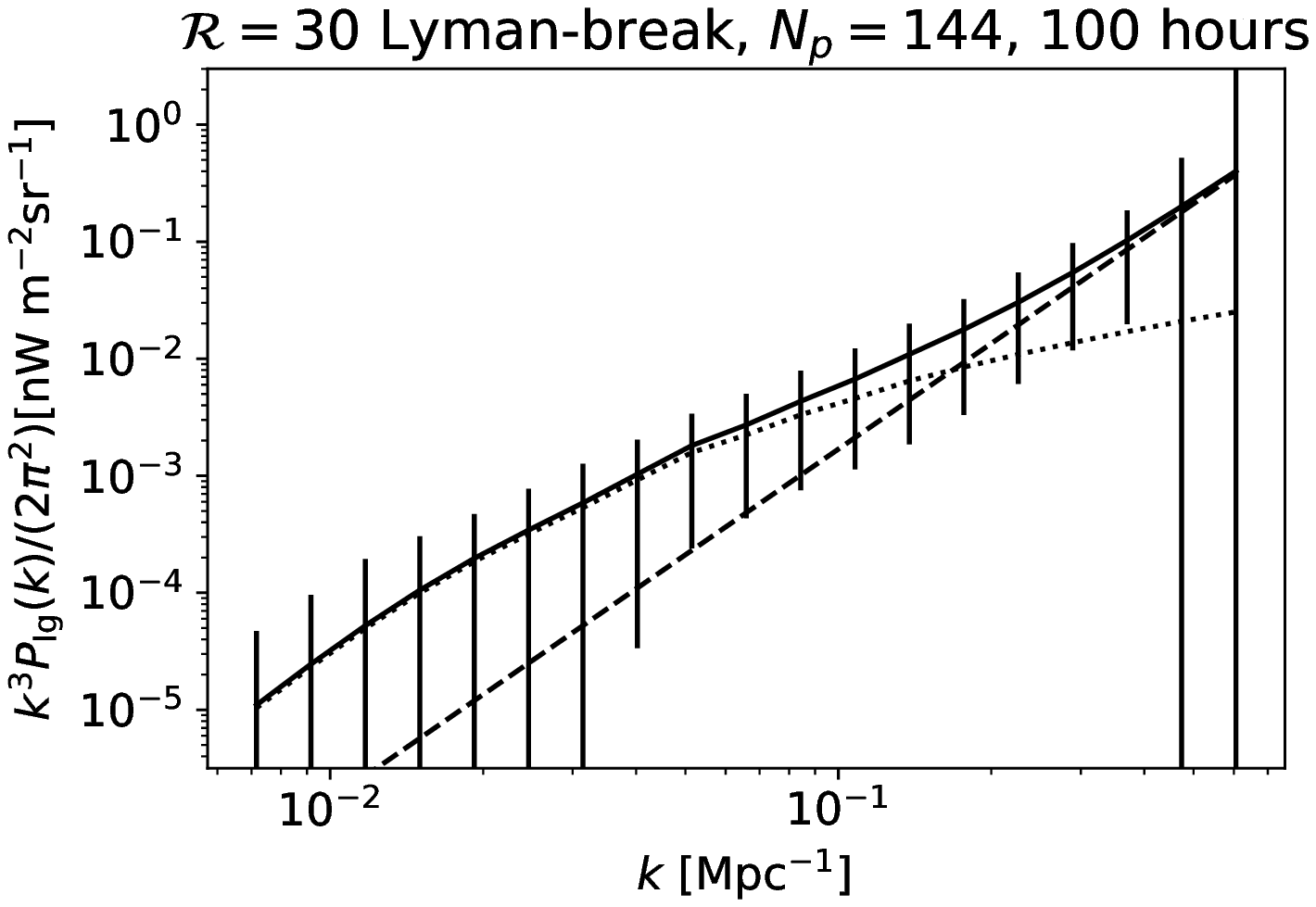}
     \includegraphics[width= 8cm]{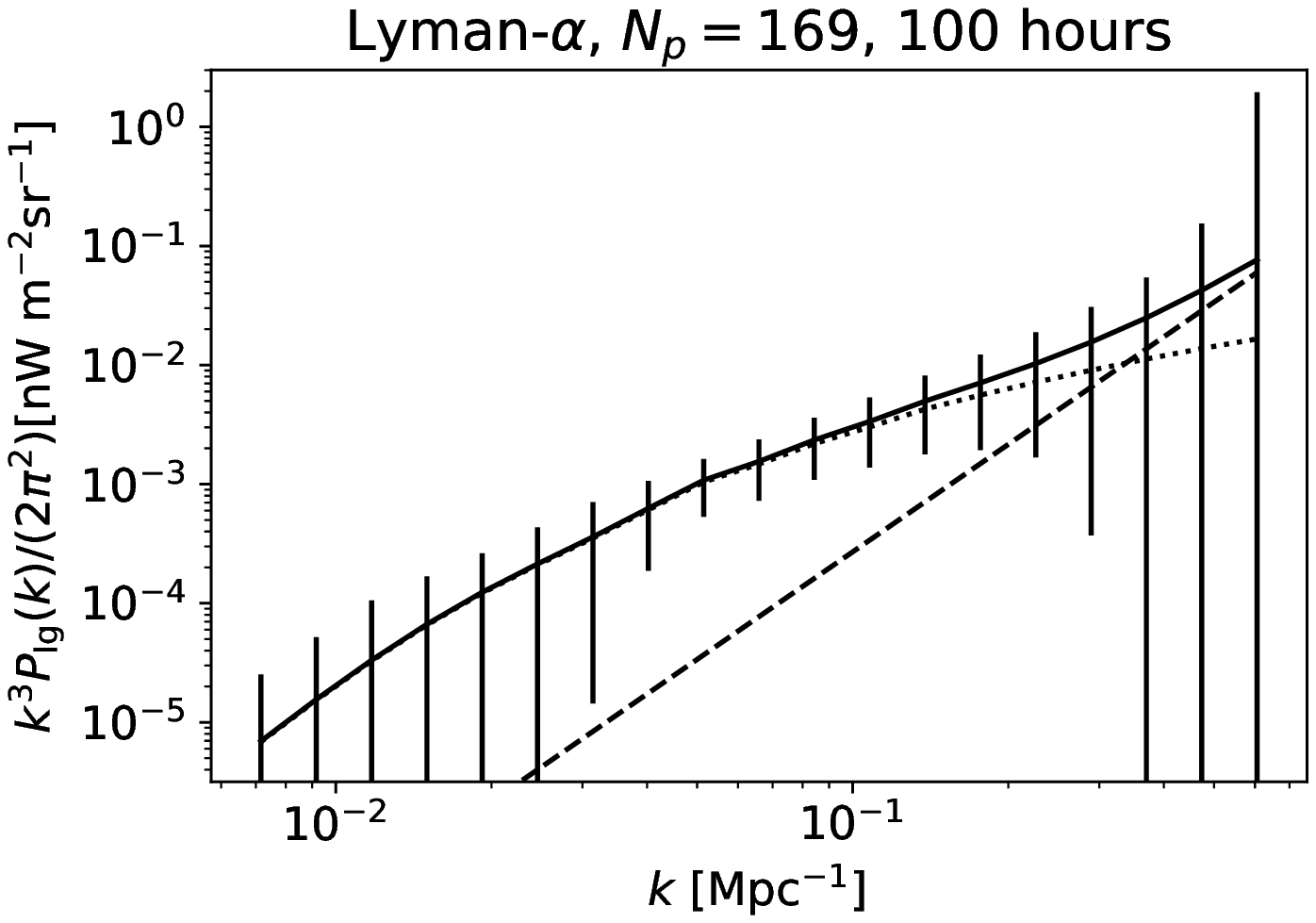}
     \caption{  \emph{Top Panels: }The total S/N for our \emph{SPHEREx} Ly$\alpha$ IM and \emph{JWST} pencil beam example as a function of the number of \emph{JWST} fields, $N_p$, for fixed total observing time.
     The left panel is for the case where \emph{JWST} galaxy redshifts are determined with the Lyman break (with ${\cal R}=30$) and the right is for the Ly$\alpha$ line.
     The solid (dotted) curves represent the S/N on the cross-power when cross-shot power is not (is) included. 
      \emph{Bottom Panels:} The galaxy-Ly$\alpha$ cross-power spectrum with errors for the Lyman-break (left panel) and Ly$\alpha$ redshift measurement case (right panel). We show errors for 100 hours of \emph{JWST} time and $N_{\rm p}=144$ ($N_{\rm p}=169$), which maximizes the S/N in the cross-power spectrum in the Lyman-break (Ly$\alpha$) case. The solid curves represent the total cross-power spectra, while the dotted and dashed represent the clustering and shot components, respectively
     }
     \label{lya_plot}
\end{figure}

In Figure \ref{lya_plot}, we show the total S/N of our Lya IM-\emph{JWST} galaxy cross correlation as a function of \emph{JWST} fields for fixed \emph{JWST} observation time (computed with Eq.~\ref{eqn:fisherApproximation}). 
We show both the conservative case without cross-shot power as well as the maximal case given by Eq.~\ref{eqn:crossShot}. 
We find that when not including cross-shot power, the sensitivity of measuring the cross correlation is approximately higher by a factor of two for \emph{JWST} redshifts measured with the Ly$\alpha$ line compared to with the Lyman-break technique (when the cross-shot power is included, the sensitivity difference is closer to a factor of $\sim$1.5). This is mainly due to the increased number of galaxies detected in the former. With 100 hours of \emph{JWST} time, a total S/N of ${\sim}5$ is found when redshifts are determined from the Ly$\alpha$ line. We find that this maximum S/N scales as the square root of the total \emph{JWST} observation time. We also note that if the H$\alpha$ interloping lines were completely removed, the S/N would increase roughly ten percent and if on the other hand their power was increased by a factor of two, the S/N would be degraded by about ten percent.
We point out that in addition to contamination from H$\alpha$ interlopers, 
 the aggregate continuum emission from foreground/background sources as well as other interloping lines must be removed. However, a detailed treatment of this
 contamination/cleaning is beyond the scope of the current work.

Because galaxy pencil-beam survey-IM cross correlations achieve higher S/N with smaller and deeper IM surveys, we have deviated from the planned \emph{SPHEREx} specifications (by assuming a smaller field integrated for a longer time). If we use the \emph{SPHEREx} deep noise adopted in Figure 2 of \cite{2022ApJ...925..136C}, our total S/N drops by roughly a factor of 4. However, given the approximate nature of our \emph{JWST} sensitivity assumptions, it may be possible to detect more galaxies than assumed here, making a reasonable S/N possible even with less optimistic \emph{SPHEREx} noise. The Ly$\alpha$ IM signal could also be higher if there is a faint previously undetected population of Ly$\alpha$ emitting galaxies not captured in our IM assumptions. Additionally, targeting a slightly lower redshift of $z=6$ improves the S/N by a factor of ${\sim}2$ (here the increase in galaxy density is somewhat counteracted by the reduced \emph{JWST} sensitivity at shorter wavelengths and lower galaxy bias). We emphasize that the main goal of this paper is to introduce the pencil-beam galaxy/IM cross correlation formalism. We defer a more precise study of the optimal S/N use cases to future work.

In the bottom panels of Figure \ref{lya_plot}, we show the cross-power spectrum with error bars, when the maximal cross-shot power (Eq.~\ref{eqn:crossShot}) is included as part of the signal. We note that in both the Lyman-break and Ly$\alpha$ galaxy detection cases, the clustering component of the cross-power spectrum dominates on large/moderate spatial scales (relative to the IM box size) and the shot component dominates on smaller spatial scales. The shot signal is generally more important in the Lyman break detection due to a larger assumed value of the duty cycle ratio appearing in Eq.~\ref{eqn:crossShot}. We note that taking a smaller number of deeper \emph{JWST} pointings, $N_{\rm p}$, increases the relative importance of the clustering versus the shot components of the cross-power spectrum. The difference is not dramatic however, because while reducing $N_{\rm p}$ reduces the shot component due to detecting fainter galaxies, it also reduces the clustering component due to lowering the galaxy bias (see Figure \ref{fig:bias_and_density}). We note that the shot power here is likely an overestimate. Thus, we expect the scales where the S/N is the highest to be dominated by the clustering power. We also note that the impact of the cross-shot power in the SVL IM example above is much smaller due to the higher value of the intensity mapping duty cycle (1 vs 0.1 in the Ly$\alpha$ IM case).

\section{21cm IM with HERA} 
\label{sec:hera}
In our final example, we explore cross correlation between \emph{JWST} pencil beams and  the \emph{HERA} 21cm survey. To compute the \emph{HERA} sensitivity, we use the default settings of the \emph{21cmSense} code \footnote{\url{https://github.com/steven-murray/21cmSense}}.  \emph{HERA} is a drift scan instrument with a $9^\circ$ wide FOV, and this code assumes the projected \emph{HERA} mission sensitivity as discussed in \citet{2013AJ....145...65P} and \citet{2014ApJ...782...66P}.  This sensitivity calculation incorporates the sensitivity loss owing to the wedge.  Because we are interested in the modes with our overlapping pencil-beam survey which we take as all within one FOV of HERA, we adapt \emph{21cmSense} to provide the per mode `imaging' sensitivity.
For our \emph{JWST} galaxies, we assume redshifts are determined with Ly$\alpha$ lines and have a spectral resolution of ${\cal R}=700$. The other \emph{JWST} observing modes described above have much lower S/N due to the wedge removing low-$k_\parallel$ Fourier modes and the high--$k_\parallel$ modes being inaccessible in the pencil beams due to poor resolution along the line-of-sight. While in principle this could be remedied by finding Lyman-break redshifts with ${\cal R}=700$, this would result in even fewer galaxy redshifts than the ${\cal R}=200$ case which is already substantially worse than Ly$\alpha$ redshifts (see Figure \ref{fig:bias_and_density}).

Following \cite{2018JCAP...10..016M}, we assume that the 21cm power spectrum is given by the perturbation-theory motivated form

\begin{equation}
    P_{21} = (20~{\rm mK})^2 \times b_1^2 (1- R_{\rm eff}^2 k^2/3)^2 P_{\rm m}(k),
\end{equation}
where we take $b_1 = -1 $ and $R_{\rm eff} = 1$\; Mpc, numbers motivated in \cite{2018JCAP...10..016M} for the last half of reionzation based on radiative transfer simulations (see their Fig. 7).  The galaxy-21cm cross-power spectrum is given by

\begin{equation}
    P_{\rm g, 21} = 20~{\rm mK} \times b_1 \bar{b}_{\rm g} (1- R_{\rm eff}^2 k^2/3) P_{\rm m}(k).
\end{equation}

\begin{figure}
     \centering
     \includegraphics[width= 8cm]{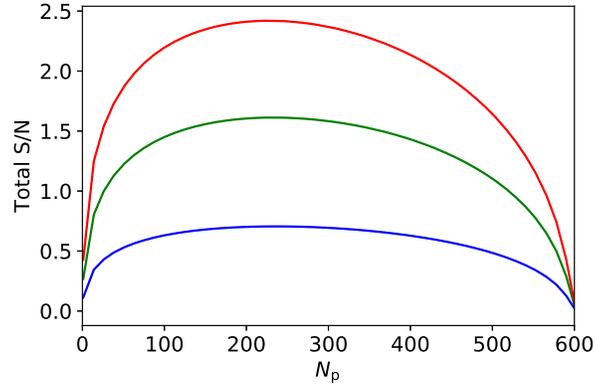}
     \caption{The total S/N on the cross-power spectrum for our \emph{HERA}-\emph{JWST} cross correlation example with 200 \emph{JWST} hours spread over $N_p$ pointings. The blue, green, and red curves are for 1x, 10x, and 100x the \emph{HERA} nominal integration time, respectively. We note the total S/N in the auto 21cm power spectrum in the single FOV of \emph{HERA} considered is 2.4, 12, and 29 for 1x, 10x, and 100x the \emph{HERA} nominal integration time, respectively.}
     \label{hera_plot}
\end{figure}

In Figure \ref{hera_plot}, the blue curves show the total S/N of the \emph{HERA}-\emph{JWST} galaxy cross correlation assuming 200 hours of total \emph{JWST} time. \emph{HERA}'s drift scan strategy is not ideal for cross correlating with narrow fields.  Instruments that point like \emph{LOFAR} and \emph{MWA} can integrate longer on a field and potentially achieve higher sensitivities.  To investigate the effect of a deeper integration, we scale down the IM noise power spectrum by a factor of 10 and 100, which could be accomplished by a 10 and 100 times longer integration time on a field, respectively.  These are shown by the green and red curves in Figure \ref{hera_plot}.
For 200 hours with \emph{JWST} with the optimal number of fields, we find a total S/N of ${\sim}0.7$, ${\sim}1.6$, and ${\sim}2.4$ for 1x, 10x, and 100x the \emph{HERA} nominal integration time, respectively.\footnote{The low S/N we find for one pointing of \emph{HERA} also suggests our 21cm signal model predicts less power than in some models.} We note that for these three different \emph{HERA} integration times the S/N on the 21cm auto-power spectrum, including only the one FOV being cross correlated would be 2.4, 12, and 29. Thus, with a high S/N detection (e.g., ${\sim} 100$) in the 21cm auto correlation, we expect that \emph{JWST} cross correlation could be used to verify that foregrounds are not strongly contaminating the signal.

\section{Discussion and Conclusions}
Line intensity mapping (IM) is a promising new technique to observe the high-redshift Universe. There are a number of ongoing and planned experiments to measure intensity maps in lines such as HI 21cm and Ly$\alpha$, as well as [CII] and CO lines. One challenge for these experiments is that astrophysical foregrounds can be orders of magnitude larger than the cosmological signal (e.g., galactic synchrotron emission in 21cm intensity maps or H$\alpha$ from lower-redshift galaxies in high-$z$ Ly$\alpha$ intensity maps). One way to ensure that residual foregrounds are not contaminating the IM signal is through cross correlation with galaxy surveys whose field overlaps. However, IM surveys typically have very large FOVs (degrees across), which are poorly matched to the small fields of view of most telescopes that can acquire coeval high-redshift galaxies (such as \emph{JWST}). Although such a mismatch prevents a deep galaxy survey over the entire IM survey, cross correlation is still possible with a galaxy survey consisting of pencil beams that cover some fraction of the IM field.

In this paper, we developed the formalism to forecast the sensitivity of cross correlations between IM and galaxy pencil-beam surveys. We utilized a Fisher matrix approach \citep[adapted from][]{2018PhRvD..98j3526V} that allowed us to compute the sensitivity of the galaxy-IM cross-power spectrum for any configuration of pencil beams within a IM survey and found a simple formula for the sensitivity of randomly positioned pencil beams that agrees well with the full Fisher calculation (at the ${\sim}1$ percent level for most $k$-bins in our examples). We found that random placement of pencil beams generally results in essentially the same total S/N as a lattice and that compact configurations, which do not span the entire IM survey, can slightly increase the S/N on small scales while sacrificing sensitivity on large scales. Configurations that span the entire IM survey are optimal. 

Using our formalism, we explored cross correlating three different IM surveys with \emph{JWST} pencil-beam galaxy surveys at $z=7$. Our exploration included: an illustrative sample variance-limited IM survey, a Ly$\alpha$ IM survey with an instrument like \emph{SPHEREx}, and lastly a 21cm survey with \emph{HERA}. Additionally, we considered three different modes of measuring redshifts with the NIRSpec multiobject spectrograph on \emph{JWST}. These included a low resolution (${\cal R} \approx 30$) observation of the Lyman break using NIRSpec/PRISM, a moderate resolution using a ${\cal R}\approx 700$ grating but binned to ${\cal R}\approx 200$, and finally a detection of Ly$\alpha$ lines again using the ${\cal R}\approx 700$ grating. We found that the Ly$\alpha$ line galaxy-identification strategy has the highest $S/N$ in cross correlation, followed by the low-resolution measurement of the Lyman break, with the high resolution of the Lyman break only faring slightly worse.  The latter two strategies produced a factor of $\sim 2$ smaller $S/N$ than the Ly$\alpha$ one. We found that unsurprisingly the total S/N is highest for a survey strategy that maximizes the number of galaxy redshifts measured across the IM survey. This is ${\sim}10$ and ${\sim}100$ pencil beams for 30 and 100 hours of total \emph{JWST} time when determining redshifts with the Ly$\alpha$ line, respectively.

In our \emph{SPHEREx}--Ly$\alpha$ example, we found that a total S/N of ${\sim}5$ can be achieved with 100 hours of \emph{JWST} time. We found that \emph{HERA} is not very well suited to cross correlation with \emph{JWST} owing to its drift-scan strategy. Telescopes capable of phasing in different directions like the future Square Kilometer Array may be better suited to such cross correlations. These correlations could be used to verify that residual foregrounds are not significant contaminants in high S/N 21cm surveys. 

Our study motivates several additional lines of inquiry. First, our formalism can be used to forecast cross correlations between \emph{any} intensity maps and \emph{any} galaxy survey that consists of many disparate pointings.  One example of the latter could be a survey performed with the \emph{Roman Space Telescope}.  However, we note that for larger FOV instruments like \emph{Roman}, it will be necessary to extend our formalism to include positions of galaxies within each individual not-so-pencil-beam field. Future work can also explore the effectiveness of this technique to cross correlate intensity maps with different lines. While HERA 21cm and SPHEREx Ly$\alpha$ are wide-field surveys, instruments targeting other lines for intensity mapping such as CO and [CII] often have much narrower fields such that again one could imagine these as a bunch of pencil beams. It could also be interesting to investigate how any planned observations for science not related to IM measurements with e.g. \emph{JWST} could be used to measure the cross correlation with IM experiments. 

\acknowledgements{We thank Jonathan Pober for assistance in computing the noise in our \emph{HERA} example. We also thank Joaquin Vieira and Kedar Phadke for useful discussions on \emph{JWST}. EV acknowledges support from NSF grant AST-2009309 and NASA ATP grant 80NSSC22K0629. MM acknowledges support from NASA 19-ATP19-0191.}

\bibliography{cc_paper}{}

\begin{thebibliography}{}
\expandafter\ifx\csname natexlab\endcsname\relax\def\natexlab#1{#1}\fi
\providecommand{\url}[1]{\href{#1}{#1}}
\providecommand{\dodoi}[1]{doi:~\href{http://doi.org/#1}{\nolinkurl{#1}}}
\providecommand{\doeprint}[1]{\href{http://ascl.net/#1}{\nolinkurl{http://ascl.net/#1}}}
\providecommand{\doarXiv}[1]{\href{https://arxiv.org/abs/#1}{\nolinkurl{https://arxiv.org/abs/#1}}}

\bibitem[{{Barry} {et~al.}(2022){Barry}, {Bernardi}, {Greig}, {Kern}, \&
  {Mertens}}]{2022JATIS...8a1007B}
{Barry}, N., {Bernardi}, G., {Greig}, B., {Kern}, N., \& {Mertens}, F. 2022,
  Journal of Astronomical Telescopes, Instruments, and Systems, 8, 011007,
  \dodoi{10.1117/1.JATIS.8.1.011007}

\bibitem[{{Beardsley} {et~al.}(2015){Beardsley}, {Morales}, {Lidz}, {Malloy},
  \& {Sutter}}]{2015ApJ...800..128B}
{Beardsley}, A.~P., {Morales}, M.~F., {Lidz}, A., {Malloy}, M., \& {Sutter},
  P.~M. 2015, \apj, 800, 128, \dodoi{10.1088/0004-637X/800/2/128}

\bibitem[{{Bouwens} {et~al.}(2015){Bouwens}, {Illingworth}, {Oesch}, {Trenti},
  {Labb{\'e}}, {Bradley}, {Carollo}, {van Dokkum}, {Gonzalez}, {Holwerda},
  {Franx}, {Spitler}, {Smit}, \& {Magee}}]{2015ApJ...803...34B}
{Bouwens}, R.~J., {Illingworth}, G.~D., {Oesch}, P.~A., {et~al.} 2015, \apj,
  803, 34, \dodoi{10.1088/0004-637X/803/1/34}

\bibitem[{{Bouwens} {et~al.}(2021){Bouwens}, {Oesch}, {Stefanon},
  {Illingworth}, {Labb{\'e}}, {Reddy}, {Atek}, {Montes}, {Naidu},
  {Nanayakkara}, {Nelson}, \& {Wilkins}}]{2021AJ....162...47B}
{Bouwens}, R.~J., {Oesch}, P.~A., {Stefanon}, M., {et~al.} 2021, \aj, 162, 47,
  \dodoi{10.3847/1538-3881/abf83e}

\bibitem[{{Carilli}(2011)}]{2011ApJ...730L..30C}
{Carilli}, C.~L. 2011, \apjl, 730, L30, \dodoi{10.1088/2041-8205/730/2/L30}

\bibitem[{{Casey} {et~al.}(2022){Casey}, {Kartaltepe}, {Drakos}, {Franco},
  {Harish}, {Paquereau}, {Ilbert}, {Rose}, {Cox}, {Nightingale}, {Robertson},
  {Silverman}, {Koekemoer}, {Massey}, {McCracken}, {Rhodes}, {Akins},
  {Amvrosiadis}, {Arango-Toro}, {Bagley}, {Bongiorno}, {Capak}, {Champagne},
  {Chartab}, {Chavez Ortiz}, {Chworowsky}, {Cooke}, {Cooper}, {Darvish},
  {Ding}, {Faisst}, {Finkelstein}, {Fujimoto}, {Gentile}, {Gillman}, {Gould},
  {Gozaliasl}, {Hayward}, {He}, {Hemmati}, {Hirschmann}, {Jahnke}, {Jin},
  {Khostovan}, {Kokorev}, {Lambrides}, {Laigle}, {Larson}, {Leung}, {Liu},
  {Liaudat}, {Long}, {Magdis}, {Mahler}, {Mainieri}, {Manning}, {Maraston},
  {Martin}, {McCleary}, {McKinney}, {McPartland}, {Mobasher}, {Pattnaik},
  {Renzini}, {Rich}, {Sanders}, {Sattari}, {Scognamiglio}, {Scoville}, {Sheth},
  {Shuntov}, {Sparre}, {Suzuki}, {Talia}, {Toft}, {Trakhtenbrot}, {Urry},
  {Valentino}, {Vanderhoof}, {Vardoulaki}, {Weaver}, {Whitaker}, {Wilkins},
  {Yang}, \& {Zavala}}]{2022arXiv221107865C}
{Casey}, C.~M., {Kartaltepe}, J.~S., {Drakos}, N.~E., {et~al.} 2022, arXiv
  e-prints, arXiv:2211.07865, \dodoi{10.48550/arXiv.2211.07865}

\bibitem[{{Cheng} \& {Chang}(2022)}]{2022ApJ...925..136C}
{Cheng}, Y.-T., \& {Chang}, T.-C. 2022, \apj, 925, 136,
  \dodoi{10.3847/1538-4357/ac3aee}

\bibitem[{{Cleary} {et~al.}(2022){Cleary}, {Borowska}, {Breysse}, {Catha},
  {Chung}, {Church}, {Dickinson}, {Eriksen}, {Foss}, {Gundersen}, {Harper},
  {Harris}, {Hobbs}, {Ihle}, {Kim}, {Kocz}, {Lamb}, {Lunde}, {Padmanabhan},
  {Pearson}, {Philip}, {Powell}, {Rasmussen}, {Readhead}, {Rennie}, {Silva},
  {Stutzer}, {Uzgil}, {Watts}, {Wehus}, {Woody}, {Basoalto}, {Bond}, {Dunne},
  {Gaier}, {Hensley}, {Keating}, {Lawrence}, {Murray}, {Paladini}, {Reeves},
  {Viero}, {Wechsler}, \& {Comap Collaboration}}]{2022ApJ...933..182C}
{Cleary}, K.~A., {Borowska}, J., {Breysse}, P.~C., {et~al.} 2022, \apj, 933,
  182, \dodoi{10.3847/1538-4357/ac63cc}

\bibitem[{{Comaschi} {et~al.}(2016){Comaschi}, {Yue}, \&
  {Ferrara}}]{2016MNRAS.463.3193C}
{Comaschi}, P., {Yue}, B., \& {Ferrara}, A. 2016, \mnras, 463, 3193,
  \dodoi{10.1093/mnras/stw2198}

\bibitem[{{CONCERTO Collaboration} {et~al.}(2020){CONCERTO Collaboration},
  {Ade}, {Aravena}, {Barria}, {Beelen}, {Benoit}, {B{\'e}thermin}, {Bounmy},
  {Bourrion}, {Bres}, {De Breuck}, {Calvo}, {Cao}, {Catalano}, {D{\'e}sert},
  {Dur{\'a}n}, {Fasano}, {Fenouillet}, {Garcia}, {Garde}, {Goupy}, {Groppi},
  {Hoarau}, {Lagache}, {Lambert}, {Leggeri}, {Levy-Bertrand},
  {Mac{\'\i}as-P{\'e}rez}, {Mani}, {Marpaud}, {Mauskopf}, {Monfardini},
  {Pisano}, {Ponthieu}, {Prieur}, {Roni}, {Roudier}, {Tourres}, \&
  {Tucker}}]{2020A&A...642A..60C}
{CONCERTO Collaboration}, {Ade}, P., {Aravena}, M., {et~al.} 2020, \aap, 642,
  A60, \dodoi{10.1051/0004-6361/202038456}

\bibitem[{{Cooray} {et~al.}(2019){Cooray}, {Chang}, {Unwin}, {Zemcov},
  {Coffey}, {Morrissey}, {Raouf}, {Lipscy}, {Shannon}, {Wu}, {Cen}, {Chary},
  {Dor{\'e}}, {Fan}, {Fazio}, {Finkelstein}, {Heneka}, {Lee}, {Linden},
  {Nayyeri}, {Rhodes}, {Sadoun}, {Silva}, {Trac}, {Wu}, \&
  {Zheng}}]{2019BAAS...51g..23C}
{Cooray}, A., {Chang}, T.-C., {Unwin}, S., {et~al.} 2019, in Bulletin of the
  American Astronomical Society, Vol.~51, 23.
\newblock \doarXiv{1903.03144}

\bibitem[{{Cox} {et~al.}(2022){Cox}, {Jacobs}, \&
  {Murray}}]{2022MNRAS.512..792C}
{Cox}, T.~A., {Jacobs}, D.~C., \& {Murray}, S.~G. 2022, \mnras, 512, 792,
  \dodoi{10.1093/mnras/stac486}

\bibitem[{DeBoer {et~al.}(2017)DeBoer, Parsons, Aguirre, Alexander, Ali,
  Beardsley, Bernardi, Bowman, Bradley, Carilli, Cheng, de~Lera~Acedo, Dillon,
  Ewall-Wice, Fadana, Fagnoni, Fritz, Furlanetto, Glendenning, Greig,
  Grobbelaar, Hazelton, Hewitt, Hickish, Jacobs, Julius, Kariseb, Kohn,
  Lekalake, Liu, Loots, MacMahon, Malan, Malgas, Maree, Martinot, Mathison,
  Matsetela, Mesinger, Morales, Neben, Patra, Pieterse, Pober, Razavi-Ghods,
  Ringuette, Robnett, Rosie, Sell, Smith, Syce, Tegmark, Thyagarajan, Williams,
  \& Zheng}]{DeBoer2017}
DeBoer, D.~R., Parsons, A.~R., Aguirre, J.~E., {et~al.} 2017, Publications of
  the Astronomical Society of the Pacific, 129, 045001,
  \dodoi{10.1088/1538-3873/129/974/045001}

\bibitem[{{Dor{\'e}} {et~al.}(2014){Dor{\'e}}, {Bock}, {Ashby}, {Capak},
  {Cooray}, {de Putter}, {Eifler}, {Flagey}, {Gong}, {Habib}, {Heitmann},
  {Hirata}, {Jeong}, {Katti}, {Korngut}, {Krause}, {Lee}, {Masters},
  {Mauskopf}, {Melnick}, {Mennesson}, {Nguyen}, {{\"O}berg}, {Pullen},
  {Raccanelli}, {Smith}, {Song}, {Tolls}, {Unwin}, {Venumadhav}, {Viero},
  {Werner}, \& {Zemcov}}]{2014arXiv1412.4872D}
{Dor{\'e}}, O., {Bock}, J., {Ashby}, M., {et~al.} 2014, arXiv e-prints,
  arXiv:1412.4872.
\newblock \doarXiv{1412.4872}

\bibitem[{{Itoh} {et~al.}(2018){Itoh}, {Ouchi}, {Zhang}, {Inoue}, {Mawatari},
  {Shibuya}, {Harikane}, {Ono}, {Kusakabe}, {Shimasaku}, {Fujimoto}, {Iwata},
  {Kajisawa}, {Kashikawa}, {Kawanomoto}, {Komiyama}, {Lee}, {Nagao}, \&
  {Taniguchi}}]{2018ApJ...867...46I}
{Itoh}, R., {Ouchi}, M., {Zhang}, H., {et~al.} 2018, \apj, 867, 46,
  \dodoi{10.3847/1538-4357/aadfe4}

\bibitem[{{Jeli\'{}c, V.} {et~al.}(2014){Jeli\'{}c, V.}, {de Bruyn, A. G.},
  {Mevius, M.}, {Abdalla, F. B.}, {Asad, K. M. B.}, {Bernardi, G.}, {Brentjens,
  M. A.}, {Bus, S.}, {Chapman, E.}, {Ciardi, B.}, {Daiboo, S.}, {Fernandez, E.
  R.}, {Ghosh, A.}, {Harker, G.}, {Jensen, H.}, {Kazemi, S.}, {Koopmans, L. V.
  E.}, {Labropoulos, P.}, {Martinez-Rubi, O.}, {Mellema, G.}, {Offringa, A.
  R.}, {Pandey, V. N.}, {Patil, A. H.}, {Thomas, R. M.}, {Vedantham, H. K.},
  {Veligatla, V.}, {Yatawatta, S.}, {Zaroubi, S.}, {Alexov, A.}, {Anderson,
  J.}, {Avruch, I. M.}, {Beck, R.}, {Bell, M. E.}, {Bentum, M. J.}, {Best, P.},
  {Bonafede, A.}, {Bregman, J.}, {Breitling, F.}, {Broderick, J.}, {Brouw, W.
  N.}, {Br\"uggen, M.}, {Butcher, H. R.}, {Conway, J. E.}, {de Gasperin, F.},
  {de Geus, E.}, {Deller, A.}, {Dettmar, R.-J.}, {Duscha, S.}, {Eisl\"offel,
  J.}, {Engels, D.}, {Falcke, H.}, {Fallows, R. A.}, {Fender, R.}, {Ferrari,
  C.}, {Frieswijk, W.}, {Garrett, M. A.}, {Grie\ss{}meier, J.}, {Gunst, A. W.},
  {Hamaker, J. P.}, {Hassall, T. E.}, {Haverkorn, M.}, {Heald, G.}, {Hessels,
  J. W. T.}, {Hoeft, M.}, {H\"orandel, J.}, {Horneffer, A.}, {van der Horst,
  A.}, {Iacobelli, M.}, {Juette, E.}, {Karastergiou, A.}, {Kondratiev, V. I.},
  {Kramer, M.}, {Kuniyoshi, M.}, {Kuper, G.}, {van Leeuwen, J.}, {Maat, P.},
  {Mann, G.}, {McKay-Bukowski, D.}, {McKean, J. P.}, {Munk, H.}, {Nelles, A.},
  {Norden, M. J.}, {Paas, H.}, {Pandey-Pommier, M.}, {Pietka, G.}, {Pizzo, R.},
  {Polatidis, A. G.}, {Reich, W.}, {R\"ottgering, H.}, {Rowlinson, A.},
  {Scaife, A. M. M.}, {Schwarz, D.}, {Serylak, M.}, {Smirnov, O.}, {Steinmetz,
  M.}, {Stewart, A.}, {Tagger, M.}, {Tang, Y.}, {Tasse, C.}, {ter Veen, S.},
  {Thoudam, S.}, {Toribio, C.}, {Vermeulen, R.}, {Vocks, C.}, {van Weeren, R.
  J.}, {Wijers, R. A. M. J.}, {Wijnholds, S. J.}, {Wucknitz, O.}, \& {Zarka,
  P.}}]{refId0}
{Jeli\'{}c, V.}, {de Bruyn, A. G.}, {Mevius, M.}, {et~al.} 2014, A\&A, 568,
  A101, \dodoi{10.1051/0004-6361/201423998}

\bibitem[{{Karoumpis} {et~al.}(2022){Karoumpis}, {Magnelli},
  {Romano-D{\'\i}az}, {Haslbauer}, \& {Bertoldi}}]{2022A&A...659A..12K}
{Karoumpis}, C., {Magnelli}, B., {Romano-D{\'\i}az}, E., {Haslbauer}, M., \&
  {Bertoldi}, F. 2022, \aap, 659, A12, \dodoi{10.1051/0004-6361/202141293}

\bibitem[{Koopmans {et~al.}(2015)Koopmans, Pritchard, Mellema, Aguirre, Ahn,
  Barkana, Bemmel, Bernardi, Bonaldi, Briggs, de~Bruyn, Chang, Chapman, Chen,
  Courty, Dayal, Ferrara, Fialkov, Fiore, \& Trott}]{koopmans}
Koopmans, L., Pritchard, J., Mellema, G., {et~al.} 2015, 001,
  \dodoi{10.22323/1.215.0001}

\bibitem[{{Kubota} {et~al.}(2018){Kubota}, {Yoshiura}, {Takahashi}, {Hasegawa},
  {Yajima}, {Ouchi}, {Pindor}, \& {Webster}}]{2018MNRAS.479.2754K}
{Kubota}, K., {Yoshiura}, S., {Takahashi}, K., {et~al.} 2018, \mnras, 479,
  2754, \dodoi{10.1093/mnras/sty1471}

\bibitem[{{La Plante} {et~al.}(2022){La Plante}, {Mirocha}, {Gorce}, {Lidz}, \&
  {Parsons}}]{2022arXiv220509770L}
{La Plante}, P., {Mirocha}, J., {Gorce}, A., {Lidz}, A., \& {Parsons}, A. 2022,
  arXiv e-prints, arXiv:2205.09770.
\newblock \doarXiv{2205.09770}

\bibitem[{{Lidz} {et~al.}(2011){Lidz}, {Furlanetto}, {Oh}, {Aguirre}, {Chang},
  {Dor{\'e}}, \& {Pritchard}}]{2011ApJ...741...70L}
{Lidz}, A., {Furlanetto}, S.~R., {Oh}, S.~P., {et~al.} 2011, \apj, 741, 70,
  \dodoi{10.1088/0004-637X/741/2/70}

\bibitem[{{Lidz} {et~al.}(2009){Lidz}, {Zahn}, {Furlanetto}, {McQuinn},
  {Hernquist}, \& {Zaldarriaga}}]{2009ApJ...690..252L}
{Lidz}, A., {Zahn}, O., {Furlanetto}, S.~R., {et~al.} 2009, \apj, 690, 252,
  \dodoi{10.1088/0004-637X/690/1/252}

\bibitem[{{McQuinn} \& {D'Aloisio}(2018)}]{2018JCAP...10..016M}
{McQuinn}, M., \& {D'Aloisio}, A. 2018, \jcap, 2018, 016,
  \dodoi{10.1088/1475-7516/2018/10/016}

\bibitem[{{Meerburg} {et~al.}(2013){Meerburg}, {Dvorkin}, \&
  {Spergel}}]{2013ApJ...779..124M}
{Meerburg}, P.~D., {Dvorkin}, C., \& {Spergel}, D.~N. 2013, \apj, 779, 124,
  \dodoi{10.1088/0004-637X/779/2/124}

\bibitem[{{Ono} {et~al.}(2013){Ono}, {Ouchi}, {Curtis-Lake}, {Schenker},
  {Ellis}, {McLure}, {Dunlop}, {Robertson}, {Koekemoer}, {Bowler}, {Rogers},
  {Schneider}, {Charlot}, {Stark}, {Shimasaku}, {Furlanetto}, \&
  {Cirasuolo}}]{2013ApJ...777..155O}
{Ono}, Y., {Ouchi}, M., {Curtis-Lake}, E., {et~al.} 2013, \apj, 777, 155,
  \dodoi{10.1088/0004-637X/777/2/155}

\bibitem[{{Ouchi} {et~al.}(2018){Ouchi}, {Harikane}, {Shibuya}, {Shimasaku},
  {Taniguchi}, {Konno}, {Kobayashi}, {Kajisawa}, {Nagao}, {Ono}, {Inoue},
  {Umemura}, {Mori}, {Hasegawa}, {Higuchi}, {Komiyama}, {Matsuda}, {Nakajima},
  {Saito}, \& {Wang}}]{2018PASJ...70S..13O}
{Ouchi}, M., {Harikane}, Y., {Shibuya}, T., {et~al.} 2018, \pasj, 70, S13,
  \dodoi{10.1093/pasj/psx074}

\bibitem[{{Planck Collaboration XVI}(2014)}]{2014A&A...571A..16P}
{Planck Collaboration XVI}. 2014, \aap, 571, A16,
  \dodoi{10.1051/0004-6361/201321591}

\bibitem[{{Pober} {et~al.}(2013){Pober}, {Parsons}, {DeBoer}, {McDonald},
  {McQuinn}, {Aguirre}, {Ali}, {Bradley}, {Chang}, \&
  {Morales}}]{2013AJ....145...65P}
{Pober}, J.~C., {Parsons}, A.~R., {DeBoer}, D.~R., {et~al.} 2013, \aj, 145, 65,
  \dodoi{10.1088/0004-6256/145/3/65}

\bibitem[{{Pober} {et~al.}(2014){Pober}, {Liu}, {Dillon}, {Aguirre}, {Bowman},
  {Bradley}, {Carilli}, {DeBoer}, {Hewitt}, {Jacobs}, {McQuinn}, {Morales},
  {Parsons}, {Tegmark}, \& {Werthimer}}]{2014ApJ...782...66P}
{Pober}, J.~C., {Liu}, A., {Dillon}, J.~S., {et~al.} 2014, \apj, 782, 66,
  \dodoi{10.1088/0004-637X/782/2/66}

\bibitem[{Pontoppidan {et~al.}(2016)Pontoppidan, Pickering, Laidler, Gilbert,
  Sontag, Slocum, Jr., Hanley, Earl, Pueyo, Ravindranath, Karakla, Robberto,
  Noriega-Crespo, \& Barker}]{10.1117/12.2231768}
Pontoppidan, K.~M., Pickering, T.~E., Laidler, V.~G., {et~al.} 2016, in
  Observatory Operations: Strategies, Processes, and Systems VI, ed. A.~B.
  Peck, R.~L. Seaman, \& C.~R. Benn, Vol. 9910, International Society for
  Optics and Photonics (SPIE), 381 -- 395, \dodoi{10.1117/12.2231768}

\bibitem[{{Pullen} {et~al.}(2014){Pullen}, {Dor{\'e}}, \&
  {Bock}}]{2014ApJ...786..111P}
{Pullen}, A.~R., {Dor{\'e}}, O., \& {Bock}, J. 2014, \apj, 786, 111,
  \dodoi{10.1088/0004-637X/786/2/111}

\bibitem[{{Renard} {et~al.}(2021){Renard}, {Gaztanaga}, {Croft}, {Cabayol},
  {Carretero}, {Eriksen}, {Fernandez}, {Garc{\'\i}a-Bellido}, {Miquel},
  {Padilla}, {Sanchez}, \& {Tallada-Cresp{\'\i}}}]{2021MNRAS.501.3883R}
{Renard}, P., {Gaztanaga}, E., {Croft}, R., {et~al.} 2021, \mnras, 501, 3883,
  \dodoi{10.1093/mnras/staa3783}

\bibitem[{{Schaerer}(2003)}]{2003A&A...397..527S}
{Schaerer}, D. 2003, \aap, 397, 527, \dodoi{10.1051/0004-6361:20021525}

\bibitem[{{Sheth} {et~al.}(2001){Sheth}, {Mo}, \&
  {Tormen}}]{2001MNRAS.323....1S}
{Sheth}, R.~K., {Mo}, H.~J., \& {Tormen}, G. 2001, \mnras, 323, 1,
  \dodoi{10.1046/j.1365-8711.2001.04006.x}

\bibitem[{{Sobacchi} {et~al.}(2016){Sobacchi}, {Mesinger}, \&
  {Greig}}]{2016MNRAS.459.2741S}
{Sobacchi}, E., {Mesinger}, A., \& {Greig}, B. 2016, \mnras, 459, 2741,
  \dodoi{10.1093/mnras/stw811}

\bibitem[{{Sun} {et~al.}(2021){Sun}, {Chang}, {Uzgil}, {Bock}, {Bradford},
  {Butler}, {Caze-Cortes}, {Cheng}, {Cooray}, {Crites}, {Hailey-Dunsheath},
  {Emerson}, {Frez}, {Hoscheit}, {Hunacek}, {Keenan}, {Li}, {Madonia},
  {Marrone}, {Moncelsi}, {Shiu}, {Trumper}, {Turner}, {Weber}, {Wei}, \&
  {Zemcov}}]{2021ApJ...915...33S}
{Sun}, G., {Chang}, T.~C., {Uzgil}, B.~D., {et~al.} 2021, \apj, 915, 33,
  \dodoi{10.3847/1538-4357/abfe62}

\bibitem[{{Tashiro} {et~al.}(2010){Tashiro}, {Aghanim}, {Langer}, {Douspis},
  {Zaroubi}, \& {Jelic}}]{2010MNRAS.402.2617T}
{Tashiro}, H., {Aghanim}, N., {Langer}, M., {et~al.} 2010, \mnras, 402, 2617,
  \dodoi{10.1111/j.1365-2966.2009.16078.x}

\bibitem[{{Tegmark} {et~al.}(1997){Tegmark}, {Taylor}, \&
  {Heavens}}]{1997ApJ...480...22T}
{Tegmark}, M., {Taylor}, A.~N., \& {Heavens}, A.~F. 1997, \apj, 480, 22,
  \dodoi{10.1086/303939}

\bibitem[{{Vanneste} {et~al.}(2018){Vanneste}, {Henrot-Versill{\'e}}, {Louis},
  \& {Tristram}}]{2018PhRvD..98j3526V}
{Vanneste}, S., {Henrot-Versill{\'e}}, S., {Louis}, T., \& {Tristram}, M. 2018,
  \prd, 98, 103526, \dodoi{10.1103/PhysRevD.98.103526}

\bibitem[{{Visbal} \& {Loeb}(2010)}]{2010JCAP...11..016V}
{Visbal}, E., \& {Loeb}, A. 2010, \jcap, 2010, 016,
  \dodoi{10.1088/1475-7516/2010/11/016}

\bibitem[{{Visbal} \& {McQuinn}(2018)}]{2018ApJ...863L...6V}
{Visbal}, E., \& {McQuinn}, M. 2018, \apjl, 863, L6,
  \dodoi{10.3847/2041-8213/aad5e6}

\bibitem[{{Vrbanec} {et~al.}(2020){Vrbanec}, {Ciardi}, {Jeli{\'c}}, {Jensen},
  {Iliev}, {Mellema}, \& {Zaroubi}}]{2020MNRAS.492.4952V}
{Vrbanec}, D., {Ciardi}, B., {Jeli{\'c}}, V., {et~al.} 2020, \mnras, 492, 4952,
  \dodoi{10.1093/mnras/staa183}

\end{thebibliography}
\bibliographystyle{aasjournal}

\begin{appendix}

\section{Fourier Conventions}
\label{ap:fourier}
\noindent Fourier Transform:
\begin{equation}
\tilde{\delta}(\bfk) = \int d^3 \bfr'  \delta(\bfr') e^{i\bfk \cdot \vec{r'}}.
\end{equation}
Inverse Fourier Transform:
\begin{equation}
\delta(\bfr) = \frac{1}{(2\pi)^3} \int d^3 \bfk'  \tilde{\delta}(\bfk') e^{-i\bfk' \cdot \vec{r}}.
\end{equation}
Partial Fourier transform in z-direction only:
\begin{equation}
\hat{\delta}(\bfx, k_\parallel) = \int dz \delta(\bfx, z) e^{i k_zz},
\end{equation}
where $\bfx$ is a 2-dimensional vector perpendicular to the line of sight and $z$ is the spatial coordinate along the line of sight.

\section{Covariance Matrices}
\label{ap:covariances}
In order to compute the covariance matrices appearing in Eq.~\ref{eqn:fullFishCross}, we need to determine the covariance between the various components of our data vectors $\bfI$ and $\bfg$. As defined in Section \ref{sec:formalism}, these components are the real and imaginary parts of the IM Fourier modes and the partially Fourier transformed galaxy overdensities in all of the pencil beams.

We begin with the correlations between IM Fourier modes. Denoting the real and imaginary parts of the Fourier modes with subscript Re and Im, for two arbitrary wavevectors indexed by $i$ and $j$ we find

\begin{equation}
\langle \tilde{{\delta}}_{\rm I, Re}(\bfk_i) \tilde{{\delta}}_{\rm I, Re}(\bfk_j) \rangle = \langle \tilde{{\delta}}_{\rm I, Im}(\bfk_i) \tilde{{\delta}}_{\rm I, Im}(\bfk_j) \rangle = \delta^K_{i,j} V P_{\rm I}(\bfk_i)/2,
\end{equation}
and

\begin{equation}
\langle \tilde{{\delta}}_{\rm I, Re}(\bfk_i) \tilde{{\delta}}_{\rm I, Im}(\bfk_j) \rangle = 0,
\end{equation}
where $\delta^K_{i,j}$ is the Kronecker delta and $V$ is the survey volume. This can be derived by expressing the modes with in terms of an amplitude and phase, $\tilde{\delta}_{\rm I} = |\tilde{\delta}|e^{i\phi}$ before taking the real/imaginary parts, $\tilde{{\delta}}_{\rm I, Re}=|\tilde{\delta}|\cos{(\phi)}$ and $\tilde{{\delta}}_{\rm I,Im}=|\tilde{\delta}|\sin{(\phi)} $. We have also utilized that fact that for a finite volume survey, we can express our definition of the power spectrum as $\langle \tilde{{\delta}}_{\rm I}(\bfk_i) \tilde{\delta}_{\rm I}(\bfk_j)^* \rangle = \delta^K_{i,j} V P_{\rm I}(\bfk_i)$ (as opposed to $\langle \tilde{{\delta}}_{\rm I}(\bfk_i) \tilde{{\delta}}_{\rm I}(\bfk_j)^* \rangle = (2\pi)^3 \delta^D(\bfk_i - \bfk_j) P_{\rm I}(\bfk_i)$ in the infinite volume case).

Next, we derive the covariance between the galaxy overdensity in our pencil-beam survey. Using the convolution theorem can write the real part of $g_i$ from Eq.~\ref{eqn:deltaG} as 

\begin{equation}
g_{i, j, \rm{Re} } = \frac{1}{(2\pi)^2} \int d^2 \bfk_\perp' \widetilde{W}_{\rm g}(\bfk_\perp') |\tilde{\delta}_{\rm g}(\bfk_\perp',k_{ \parallel,i})| \cos (\phi' -\bfk_\perp' \cdot \bfx_{j}),
\end{equation}
where $i$ is the index for the component of the wavenumber along the line-of-sight and $j$ is the index of the pencil beam with FOV centered on $\bfx_j$. Here $\phi'$ is the phase of the mode at $\bfk '$. Note that the imaginary part, $g_{i,j, \rm{Im}}$ is the same, but with cosine switched to sine. By correlating this equation and the real/imaginary parts of the IM Fourier modes expressed in terms of the amplitude and phase and simplifying with trigonometric identities, we derive relatively simple formulae for all of the remaining elements which appear in our covariance matrices.

The cross-correlations between the pencil-beam galaxy overdensities are then given by

\vspace{3mm}

\begin{equation}
\langle g_{i, a, \rm{Re}} ~ g_{j, b, \rm{Re}} \rangle = \langle g_{i, a, \rm{Im}} ~ g_{j, b, \rm{Im}} \rangle =  
\frac{d_\parallel \delta^K_{i,j}} {2(2\pi)^2 }   \int d^2 \bfk_\perp' P_{\rm  g}(\bfk_\perp',k_{ \parallel,i})\tilde{W}^2_{\rm g}(\bfk_\perp') \cos(\bfk_\perp' \cdot (\bfx_{\rm a} - \bfx_{\rm b})),
\end{equation}
and 
\begin{equation}
\langle g_{i, a, \rm{Re}} ~ g_{j, b, \rm{Im}} \rangle  =  0,
\end{equation}
where $d_\parallel$ is the length of the survey along the light on sight. We solve this integral numerically using Fast Fourier transforms to compute the covariance matrices in the calculations above. 

Similarly, we derive the cross-correlation terms between IM Fourier modes and pencil-beam galaxy overdensities. These are given by

\begin{equation}
 \langle g_{ i, a,{\rm Re }} ~ \tilde{{\delta}}_{\rm I, Re}(\bfk_\perp,k_{\parallel,j}) \rangle = 
  \langle g_{ i, a,{\rm Im} } ~ \tilde{{\delta}}_{\rm I, Im}(\bfk_\perp,k_{\parallel,j}) \rangle = 
\frac{1}{2} P_{\rm Ig} \left (\bfk_\perp,k_{\parallel,i} \right ) \widetilde{W}_{\rm g}(\bfk_\perp) \cos(\bfk_{\perp} \cdot \bfx_{\rm a}) d_\parallel \delta^{\rm K}_{ij}, 
\end{equation}
 
 \begin{equation}
  \langle g_{ i, a,{\rm Re} } ~ \tilde{{\delta}}_{\rm I, Im}(\bfk_\perp,k_{\parallel,j}) \rangle = 
\frac{1}{2} P_{\rm  Ig} \left (\bfk_\perp,k_{\parallel,i} \right ) \widetilde{W}_{\rm g}(\bfk_\perp) \sin(\bfk_{\perp} \cdot \bfx_{\rm a}) d_\parallel \delta^{\rm K}_{ij}, 
\end{equation}
 
 and

 \begin{equation}
  \langle g_{i, {\rm a, Im} } ~ \tilde{{\delta}}_{\rm I, Re}(\bfk_\perp,k_{\parallel,j}) \rangle = 
-\frac{1}{2} P_{\rm Ig} \left (\bfk_\perp,k_{\parallel,i} \right ) \widetilde{W}_{\rm g}(\bfk_\perp) \sin(\bfk_{\perp} \cdot \bfx_{\rm a}) d_\parallel \delta^{\rm K}_{ij}.
\end{equation}
 
 While we have worked with purely real quantities in our covariance matrices to simplify the required numerical computations, we note that very similar equations can be derived with complex data vectors.

\section{Minimum variance cross power estimator}
\label{sec:crossestim}

We want to construct the minimum variance quadratic estimator for cross correlations.  We follow the calculation in \citet{2018PhRvD..98j3526V} for the CMB angular cross power spectra, generalizing their derivation to arbitrary cross correlations and to estimate arbitrary parameters.  Our cross power covariance can be written as 

\begin{equation}
\bfC_{\rm gI} = \langle \bfg  \bfdelta_{\rm I}^T \rangle = \bfC_{\rm gI, 0} +  \Rmat_j \, \delta p_j,  
\end{equation}
where $\bfg$ and $\boldsymbol{\delta}_{\rm I}$ are vectors for galaxy and intensity mapping data sets defined in the previous appendix. The last line uses that near some reference value for the parameter $[ p_j ]_{0}$ we can approximate the covariance as linear in the parameter $p_j$, with $\Rmat_j \equiv \partial \vec{C}_{\rm gI}/\partial p_j$ evaluated at ${[ p_j ]_{0}}$ and defining $ \delta p_j \equiv  {p}_j - [ p_j ]_{0}$. 
 
A general estimator that is quadratic in our two data sets is given by

\begin{equation}
    \widehat{y}_j = {\bfg }^T \bfE_j \bfdelta_{\rm I} - b_j,
\end{equation}
where $b_j = {\rm Tr}[ \vec{E}_j \vec{C}_{\rm gI,0}^T]$, as this yields an unbiased estimator.  The expectation value of the estimator is given by $\langle \widehat{y}_j \rangle = {\rm Tr}[ \vec{E}_j \vec{C}_{\rm gI}^T] -b_j = {\rm Tr}[\vec{E}_j \vec{R}_i^T]\delta p_i$.  To create an unbiased estimator for the $\delta p_i$, we take linear combinations of the $\widehat{y}_j$:
\begin{eqnarray}
\widehat{\delta p_i} = [\bfW^{-1}]_{ij} \widehat{y}_j~~~~~\text{where}~~~~W_{ij} \equiv {\rm Tr}[\vec{E}_i \vec{R}_j^T],
\end{eqnarray}
where we show the parameter indices explicit and the measurement pixel indices implicitly as matrix multiplications.  Repeated explicit indices indicate summation.

We can now compute the estimator covariance assuming Gaussianity:
\begin{eqnarray}
    {\rm Cov} [ \widehat{\delta p_l} \widehat{\delta p_m}] \rangle &=& [{\bfW}^{-1}]_{li} [{\bfW}^{-1}]_{mj}     ~{\rm Cov} [ \widehat{y_i} \widehat{y_j} ]; \\
   {\rm Cov} [ \widehat{y_i} \widehat{y_j} ] &=&   {\rm Tr} \left[\vec{C}_{{\rm gg}}\vec{E}_i \vec{C}_{\rm II} \vec{E}_j^T \right]  +  {\rm Tr} \left[\vec{C}_{\rm gI}\vec{E}_i^T \vec{C}_{\rm gI}\vec{E}_j^T\right].
\end{eqnarray}

We want to minimize the variance in the $\widehat{\delta p_l}$ to find their optimal estimators.  The minimization of the variance of the $N_p$ estimators, $\langle \widehat{\delta p_l} \widehat{\delta p_l} \rangle$, can be performed with gradient descent-like algorithms.  
However, to make traction analytically, we make the standard approximation that if we minimize the variance of each of the $\widehat{y_l}$ individually, the estimator that results will be near the minimum for the $\widehat{\delta p_j}$.\footnote{This is the approach that leads to the standard Fisher matrix expression for auto-power, namely $F_{ij} = {\rm Tr}[\bfC^{-1}\partial \bfC/ \partial {p_i} \bfC^{-1} \partial \bfC/ \partial {p_j}]$.}  This is partly motivated by the expectation that, if our parameters are power spectrum bandpowers, we expect our estimators $y_j$ are largely diagonal as different modes are weakly correlated. 

Thus, we aim now to minimize the variance of the $\widehat{y}_i$ and to avoid the trivial solution $\vec{E}_j = 0$ we further impose the constraint that the diagonals of our weighting kernel ${W}_{jj} \equiv {\rm Tr}[\vec{E}_i \vec{R}_i^T]$ are finite by adding a Lagrange multiplier (as otherwise the minimum variance estimator that returns $\widehat{y}_i = 0$ would be selected!).  Thus, the minimum variance estimator is the derivative with respect to $\vec{E}_i$ of
\begin{eqnarray}
\langle \widehat{y_i}^2 \rangle - 2 \lambda  \left( {\rm Tr}[\vec{E}_i \vec{R}_i^T] - \beta_i \right),
\end{eqnarray}
where the $\beta_i$ are unspecified constants.  This yields
\begin{eqnarray}
\vec{C}_{{\rm gg}}\vec{E}_i \vec{C}_{\rm II}  + \vec{C}_{\rm gI} \vec{E}_i^T \vec{C}_{\rm gI} = \lambda \vec{R}_i,
\end{eqnarray}
where we have used repeatedly the identity $\partial_{\vec{A}} \text{Tr}[\vec{A}\vec{B}] = \vec{B}^T$ (c.f.~\citealt{2018PhRvD..98j3526V} for more details).   An approximate solution can be found in the applicable limit of when the noise in the auto-power dominates:
\begin{eqnarray}
\vec{C}_{{\rm gg}}\vec{E}_{i} \vec{C}_{\rm II}   = \lambda \vec{R}_i.
\label{eqn:highnoise}
\end{eqnarray}
However, the low-noise limit in which both the g and I fields are noiseless biased tracers of the same field also yields the same estimator \citep{2018PhRvD..98j3526V}, suggesting the estimator may be nearly optimal even beyond the high noise-limit that is assumed. We are free to choose $\lambda = 1/2$ since our estimate $\widehat{\delta p}_i$ as this yields an unbiased estimator (\citet{2018PhRvD..98j3526V}; when g=I this gives the standard Fisher Matrix expression).

Solving either Eq. \ref{eqn:highnoise} for $\vec{E}_i$ yields

\begin{equation}
\widehat{\vec{E}}_i \approx \frac{1}{2} \vec{C}_{{\rm gg}}^{-1} \vec{R}_i \vec{C}_{\rm II}^{-1},
\end{equation}
and
\begin{equation}
W_{ij} = \frac{1}{2}  {\rm Tr}[\vec{C}_{{\rm gg}}^{-1} \vec{R}_i \vec{C}_{\rm II}^{-1}\vec{R}_j^T],
\label{eqn:Wij}
\end{equation}
such that the variance is
\begin{eqnarray}
\langle \widehat{\delta p_l} \widehat{\delta p_m} \rangle &=&   \frac{1}{2}  [{W}^{-1}]_{ki} [{W}^{-1}]_{lj}\left(W_{ij} + G_{ij} \right); \label{eqn:finalvar}\\
%G_{ij} &=&  {\rm Tr} \Big[\vec{C}_{\rm gI}\vec{C}_{\rm II}^{-1} \vec{R}_i^T \vec{C}_{{\rm gg}}^{-1}\vec{C}_{\rm gI}\vec{C}_{\rm II}^{-1} \vec{R}_j^T \vec{C}_{{\rm gg}}^{-1}\Big].\\
G_{ij} &=&  \frac{1}{2}  {\rm Tr} \Big[  \vec{C}_{{\rm gg}}^{-1} \vec{R}_i \vec{C}_{\rm II}^{-1}\vec{C}_{\rm gI}^T \vec{C}_{{\rm gg}}^{-1}\vec{R}_j  \vec{C}_{\rm II}^{-1} \vec{C}_{\rm gI}^T \Big].
\label{eqn:Gij}
\end{eqnarray}
In the high noise, the second term in Eq.~\ref{eqn:finalvar} (with $\vec{G}$) is more important. 
When shot noise dominates, we can treat $\vec{C}_{{\rm gg}}$ as diagonal, and  $\vec{C}_{\rm II}$ is diagonal if we choose the Fourier basis.  {\it Unlike the main body of the paper, where we consider real and imaginary components of modes, for notational simplicity we deal with complex modes; we remark at the end of this derivation how the final formulas relate to those in the main body.} 

In this paper, $\delta p_i$ are the bandpowers of the cross such that $p_l \rightarrow P_{\rm Ig}(\vec{k}_\perp, k_\parallel)$.  Let us first treat the case where only a single mode contributes to the band power estimate.  Then working at fixed $k_{\parallel}$, since we can treat each independently, our matrices become 
\begin{eqnarray}
[\vec{C}_{{\rm gg}}]_{ij} &\approx & 2 \sigma_{\rm g}(k_{\parallel})^{2} \delta^{\rm K}_{ij}~~  \overbrace{\Longrightarrow}^{\rm shot}~~ {\frac{d_\parallel}{L^2 \bar{n}_{\rm g}}  \delta^{\rm K}_{ij}}, \label{eqn:CAA}\\
{[\vec{C}_{\rm II}]}_{\vec{k}_\perp \vec{k}_\perp'} &=& {[V P_{\rm I}( \vec{k}_{\perp}, k_{\parallel})]} \; \delta^{\rm K}_{\vec{k}_{\perp} -\vec{k}_{\perp}'},\\
{[\vec{C}_{\rm gI}]}_{{\rm A} \vec{k}_\perp } &=& \exp[-i \vec{k}_\perp \cdot \vec{x}_{{\rm A}}] P_{\rm Ig}( \vec{k}_{\perp}, k_{\parallel}) d_\parallel,\\
{[\vec{R}]}_{ {\rm A} \vec{k}_\perp} &\equiv& \frac{\partial {[\vec{C}_{\rm gI}]}_{{\rm A} \vec{k}_\perp }}{\partial P_{\rm Ig}( \vec{k}_{\perp}, k_{\parallel})}  =  \exp[-i \vec{k}_\perp \cdot \vec{x}_{{\rm A}}] d_\parallel. 
\end{eqnarray}
where $2 \sigma_{\rm g}^2 = \langle |g_{i, A}|^2\rangle/2 = \langle {\rm Re}[g_{i, A}]^2\rangle$. In the top line, to evaluate in the limit shot noise dominates, we used that $2 \sigma_{\rm g}^2 = d_\parallel (L^2 \bar{n}_{\rm g})^{-1}$, where $\bar{n}_{\rm g}$ is the 3D number density of galaxies and $L$ is the transverse size of each (square) JWST field.   For notational simplicity, we do not include the field-of-view window functions $\widetilde{W}_g(\vec{k}_\perp)$ that accompany all the terms (one for every $g$ subscript in the covariances) associated with the galaxy survey.  We put these terms back in in the main text, although they are unity for modes with wavelengths much larger than the survey field. 

With these choices and simplifications, Eqns.~\ref{eqn:Wij} and \ref{eqn:Gij} become
\begin{eqnarray}
    2 W_{{\vec{k}_{\perp, i}}, \vec{k}_{\perp, j}} &=& [V\ P_{\rm I}]^{-1} \ (2\sigma_{\rm g}^2)^{-1} \ N_{\rm g} \ d_\parallel^2 \ \delta^{\rm K}_{{\vec{k}_{\perp, i}}, \vec{k}_{\perp, j}}  \approx  \overbrace{P_{\rm I}^{-1} \bar{n}_{\rm g} f_{\rm cov}  \delta^{\rm K}_{{\vec{k}_{\perp, i}}, \vec{k}_{\perp, j}}}^{\rm shot~ noise~limit}; \\\
    2 G_{{\vec{k}_{\perp, i}}, \vec{k}_{\perp, j}} &=&  (2\sigma_{\rm g}^2)^{-2} d_\parallel^4 [V P_{\rm I}(\vec{k}_{\perp, i})]^{-1} [V P_{\rm I}(\vec{k}_{\perp, j})]^{-1}P_{\rm Ig}(\vec{k}_{\perp, i})P_{\rm Ig}(\vec{k}_{\perp, j}) \sum_{\rm \forall~\vec{\Delta x}}\exp[-i \vec{\Delta x} \cdot (\vec{k}_{\perp, i} - \vec{k}_{\perp, j})],\\ %\delta^{\rm K}_{{\vec{k}_{\perp, i}}, \vec{k}_{\perp, j}}%\approx  \overbrace{P_{\rm I}^{-2} P_{\rm Ig}^2 \bar{n}_{\rm g}^2 f_{\rm cov}^2  \delta^{\rm K}_{{\vec{k}_{\perp, i}}, \vec{k}_{\perp, j}}}^{\rm +shot~ noise~limit}
    &\approx & (2\sigma_{\rm g}^2)^{-2} d_\parallel^4 N_{\rm g}^2 [V P_{\rm I}(\vec{k}_{\perp, i}, k_\parallel)]^{-2} P_{\rm Ig}(\vec{k}_{\perp, i}, k_\parallel)^2 \delta^{\rm K}_{{\vec{k}_{\perp, i}}, \vec{k}_{\perp, j}}, \\
    & \underbrace{\Longrightarrow}_{\rm shot}&  f_{\rm cov}^2 \bar \bar{n}_{\rm g}^{2}  [P_{\rm I}(\vec{k}_{\perp, i}, k_\parallel)]^{-2} P_{\rm Ig}(\vec{k}_{\perp, i}, k_\parallel)^2 \delta^{\rm K}_{{\vec{k}_{\perp, i}}, \vec{k}_{\perp, j}},
\end{eqnarray}
where $N_{\rm g}$ is the number of pointings and $f_{\rm cov} = N_{\rm g} L^2/(d_x d_y)$ is the covering fraction of pointings.   The sum over all $\vec{\Delta x}$ is over all pairs of pencil-beam pointings, and the second to last line uses the approximation $\sum_{\rm \forall~ \vec{\Delta x}}\exp[-i \vec{\Delta x} \cdot (\vec{k}_{\perp, i} - \vec{k}_{\perp, j})] \approx N_{\rm g} \delta^{\rm K}_{{\vec{k}_{\perp, i}}, \vec{k}_{\perp, j}}$.

With these simplifications, our estimate for the error on the cross power is

\begin{equation}
\langle \delta P_{\rm Ig}^2 \rangle =    V \,  P_{\rm I}(\vec{k}_\perp, k_\parallel) \ (2\sigma_{\rm g}^2)\, d_\parallel^{-2} N_{\rm g}^{-1} +  P_{\rm Ig}(\vec{k}_\perp, k_\parallel)^2,
\end{equation}
or assuming shot noise dominates the variance in each JWST pointing:

\begin{equation}
\langle \delta P_{\rm Ig}^2 \rangle =   P_{\rm I}(\vec{k}_\perp, k_\parallel) \ (f_{\rm cov} \ \bar{n}_{\rm g})^{-1}   +  P_{\rm Ig}(\vec{k}_\perp, k_\parallel)^2,
\label{eqn:varshot}
\end{equation}
a simple form we might have guessed without all of this work.  To generalize to multiple modes in a bandpower bin, error should be summed in inverse quadrature, which, in the limit of isotropic power spectrum, divides by square root the number of modes in a bandpower.  Because we are dealing with complex modes, this counts both the real and imaginary component, in contrast to the main body of the paper where we only count the half plane of modes and write single mode expressions from the combined constraint of both the real and imaginary components (leading to a factor of two smaller variances per mode but half the total modes).  This reconciles the factor of two differences between the expressions here and in \S~\ref{sec:formalism}, and we include the window function terms there that we have omitted in the main text.

This estimate for the error includes sample variance uncertainty, which is the second $P_{\rm Ig}^2$ term in the previous two equations.  This term means that the S/N on a mode can never be greater than unity. However, we do not care about sample variance when asking how well cross correlations can be detected, it is only the noise on the mode that matters, and so in principal the S/N can be arbitrarily large.  In practice this distinction is not so important, as most of our modes are noise dominated.  The next section rederives the error in a manner that does not include sample variance.

\section{Signal-to-noise estimate without sample variance}
\label{sec:signtonoisedetection}

We wish to calculate the S/N ratio without sample variance.  The S/N ratio that cross correlations can be detected assuming Gaussian noise and not including sample variance in the noise is
\begin{equation}
({\rm S/N})^2 =  (g_{i}  \delta_{{\rm I}j} )  [\bfC^{\rm N}]^{-1}_{(ij) (k l)} (g_{k} \delta_{{\rm I}l} ),
\label{eqn:Sigtonoise}
\end{equation}
where all indexes are summed and the covariance matrix is given by
$$
[\bfC^{\rm N}]_{(ij) (k l)} = \langle g_{i}  \delta_{{\rm I}j}  g_{k} \delta_{{\rm I}l}\rangle - \langle g_{i}  \delta_{{\rm I}j} \rangle \langle g_{k} \delta_{{\rm I}l}\rangle
$$
where  $g_{i}$ indicates the $i^{\rm th}$ pointing, and \emph{in this section only} the brackets only ensemble average over the noise and not over realizations of the galaxy and intensity mapping fields.   We group $(ij)$ and $(kl)$ in equation~\ref{eqn:Sigtonoise}, as each distinct integer pair should be considered as one entry in the noise covariance matrix.

To compute the covariance matrix, we do not want to ensemble average over different realizations of the galaxy and intensity mapping fields.  Rather, since we are interested in how well cross correlations can be detected for a given intensity mapping and galaxy field, we only want to average over the noise.  Let us split up a field into both its signal and noise so  $g_i = \langle g_i \rangle + g_i^{\rm N}$ and $ \delta_{{\rm I}j} = \langle  \delta_{{\rm I}j} \rangle +  \delta_{{\rm I}j}^{\rm N}$, where since we are not ensemble averaging over pixels or modes the $\langle g_i \rangle$ and $\langle  \delta_{{\rm I}j} \rangle$ are particular values.  Since we assume the different overdensity fields have uncorrelated noise $\langle g_{i}  \delta_{{\rm I}j} \rangle = \langle g_i \rangle \langle \delta_{{\rm I}j} \rangle$,  and thus 
\begin{eqnarray}
[\bfC^{\rm N}]_{(ij) (k l)} &=&   \left\langle \left( \langle g_i \rangle + g_i^{\rm N} \right) \left( \langle  \delta_{{\rm I}j} \rangle +  \delta_{{\rm I}j}^{\rm N} \right) \left( \langle g_k \rangle + g_k^{\rm N} \right) \left( \langle  \delta_{{\rm I}l} \rangle +  \delta_{{\rm I}l}^{\rm N} \right)  \right \rangle - \langle g_i \rangle \langle \delta_{{\rm I}j} \rangle \langle g_k \rangle \langle \delta_{{\rm I}l} \rangle  \\
&=&  \langle g_i \rangle \langle g_k \rangle C_{{\rm II}jl}^{\rm N} +  C_{{\rm gg}ik}^{\rm N} \langle  \delta_{{\rm I}j} \rangle \langle  \delta_{{\rm I}l} \rangle  +     C_{{\rm gg}ik}^{\rm N} C_{{\rm II}jl}^{\rm N}
\end{eqnarray}
where $C_{{\rm gg}}^{\rm N}$ is the uncorrelated shot noise and $C_{{\rm II}}^{\rm N}$ the instrument noise, although some  other stochastic uncorrelated cosmological part could also come into either.  While shot noise could correlate, the component of shot noise that does correlate we treat as signal. % Both noise terms are diagonal in the models used in this paper and so inversion is easy.

To calculate a typical S/N, we just then substitute the expectation values
$$
\langle {\rm S/N} \rangle^2 =  \langle g_{i}  \delta_{{\rm I}j}\rangle   [\bfC^{\rm N}]^{-1}_{(ij) (k l)} \langle g_{k} \delta_{{\rm I}l}\rangle = C_{{\rm gI}ij}  [\bfC^{\rm N}]^{-1}_{(ij) (k l)}C_{{\rm gI}kl},
$$
where $\langle {\rm S/N} \rangle$ is our notation for the `typical S/N', we used the definitions  $C_{{\rm gI}ij} = \langle g_{i}  \delta_{{\rm I}j}\rangle$, and we mean to evaluate the $[\bfC^{\rm N}]^{-1}_{(ij) (k l)} $ with typical values for $\langle g_i \rangle$ and $\langle  \delta_{{\rm I}j} \rangle $.

The Fisher matrix is defined as the curvature of minus the log likelihood (since we have assumed Gaussianity):
\begin{equation}
\bfF_{p_a p_b} \equiv  \partial_{p_1} \partial_{p_2}   (g_{i}  \delta_{{\rm I}j}-C_{{\rm gI}ij}(\boldsymbol{p}) )  [2 \bfC^{\rm N}]_{(ij)^{-1} (k l)} (g_{k} \delta_{{\rm I}l}-C_{{\rm gI}kl}(\boldsymbol{p})  )  = \partial_{p_1} \partial_{p_2}  \langle {\rm S/N} \rangle^2/2,
\end{equation}
where here the $C_{{\rm gI}ij}(\boldsymbol{p})$ are the model covariance matrices where $\boldsymbol{p}$ is the vector of parameters and the $g_{i}  \delta_{{\rm I}j}$ is the product of the measured `pixels' in each survey. (The parameters in this case, since we are not constraining a statistical theory in contrast to the previous section, are likely to be the power in each wavenumber bin.) Let us specialize to the case where the parameters are the power spectrum of a given mode $P_{\rm Ig}(\bfk_i)$:

\begin{eqnarray}
F_{\bfk_i \bfk_j} &=&  {\rm Tr} \left[ \frac{\partial \bfC_{\rm gI}}{\partial P_{\rm Ig}(\bfk_i)} [\bfC^{\rm N}]^{-1} \frac{\partial \bfC_{\rm gI}}{\partial P_{\rm Ig}(\bfk_j)}\right];\\
           &=& {\rm Tr} \left[ \bfR [\bfC^{\rm N}]^{-1} \bfR\right],
\end{eqnarray}
where $\bfR_{{\rm A} \bfk} = \exp[-i \vec{k}_\perp \cdot \vec{x}_{{\rm A}}] \, d_\parallel$ as defined in Appendix~\ref{sec:crossestim}  (also here omitting the $\widetilde{W}_g(\vec{k}_\perp)$ terms that we put back in the main text).  We use the noise-only generalization of our previous results that $C_{{\rm gg}ij}^{\rm N} =  2 [\sigma_{\rm g}^{\rm N}]^2 \delta_{ij}^{\rm K}$  and $C_{{\rm II}ij}^{\rm N} = V P_{\rm I}^{\rm N}(\bfk_i) \delta_{\bfk_i, -\bfk_j}^{\rm K}$, where superscript N indicates noise-only. Furthermore,  $\sum_{\rm \forall~ \vec{\Delta x}}\exp[-i \vec{\Delta x} \cdot (\vec{k}_{\perp, i} - \vec{k}_{\perp, j})] \approx N_{\rm g} \delta^{\rm K}_{{\vec{k}_{\perp, i}}, \vec{k}_{\perp, j}}$.  We also need a mode amplitude to compute $\bfC^{\rm N}$, which we take to be the average 
$\langle g_i \rangle \langle g_j \rangle  = 2 [\sigma_{\rm g}^{\rm N}]^2 \delta_{ij}^{\rm K}$  and  $\langle  \delta_{{\rm I}}(\bfk_i) \delta_{{\rm I}}(\bfk_j)  \rangle = V P_{\rm I} (\bfk_i) \delta_{\bfk_i, -\bfk_j}^{\rm K}$.  This is an approximation that ignores the cross power of the modes and is strictly valid when the noise dominates over the signal. With this approximation, the Fisher matrix becomes
%\begin{equation}
%   F_{\bfk_i \bfk_j} =  f_{\rm cov} \bar{n}_{\rm g} [P^{\rm N}_{\rm I}(\bfk_i)]^{-1} \delta_{ij}^{\rm K}.
%\end{equation}
\begin{equation}
  F_{\bfk_i \bfk_j} \approx      \left( P_{\rm I}(\vec{k}_i) \ (2\sigma_{\rm g}^2) - P_{\rm I}^{\rm SV}(\vec{k}_i) \ (2[\sigma^{\rm SV}_{\rm g}]^2) \right)^{-1} \, d_\parallel^{2} N_{\rm g}/V ~\delta_{\bfk_i \bfk_j}^{\rm K},
\end{equation}
where we note that $P_{\rm I} = P_{\rm I}^{\rm SV} + P_{\rm I}^{\rm N}$ and similarly $\sigma_{\rm g}^2 = [\sigma_{\rm g}^{\rm SV}]^2 + [\sigma_{\rm g}^{\rm N}]^2$.
Noting that $\langle \delta P_{\rm Ig}(\bfk_i)^2 \rangle  = F_{\bfk_i \bfk_i}^{-1}$ this matches this limit of our previous derivation with the sample variance term subtracted off (c.f. Eq.~\ref{eqn:varshot}).  Putting back the window function terms this yields equation~\ref{eqn:noSampleApproximation} in the main text.

\label{ap:estimator}

\end{appendix}

\end{document}